\documentclass[preprint,12pt]{elsarticle}

\usepackage{subcaption}
\usepackage{graphicx}
\usepackage{float}
\usepackage{booktabs} 
\usepackage{multirow} 

\usepackage{amssymb}
\usepackage{amsmath}
\usepackage{longtable}
\usepackage{changepage}
\usepackage{geometry}
\usepackage{booktabs}
\usepackage{hyperref}
\usepackage{color}

\journal{XXXXXXX}

\begin{document}

\begin{frontmatter}



\title{Urban Forms Across Continents: A Data-Driven Comparison of Lausanne and Philadelphia}

\author[label1]{Arthur Carmès}
\author[label1]{Léo Catteau}
\author[label1]{Andrew Sonta}
\ead{andrew.sonta@epfl.ch}
\author[label2]{Arash Tavakoli}
\ead{arash.tavakoli@villanova.edu}

\affiliation[label1]{organization={Engineering and Technology for Human-Oriented Sustainability (ETHOS) Lab, Ecole Polytechnique Federal de Lausanne (EPFL)},
            addressline={HBL 1 3B (Halle Bleue)},
            city={Fribourg},
            postcode={1700},
            country={Switzerland}}

\affiliation[label2]{organization={Human Centered Cities Lab, Villanova University},
            addressline={800 E Lancaster Ave},
            city={Villanova},
            postcode={19085},
            state={PA},
            country={USA}}

\begin{abstract}
Understanding urban form is crucial for sustainable urban planning and enhancing quality of life. This study presents a data-driven framework to systematically identify and compare urban typologies across geographically and culturally distinct cities. Using open-source geospatial data from OpenStreetMap, we extracted multidimensional features related to topography, multimodality, green spaces, and points of interest for the cities of Lausanne, Switzerland, and Philadelphia, USA. A grid-based approach was used to divide each city into Basic Spatial Units (BSU), and Gaussian Mixture Models (GMM) were applied to cluster BSUs based on their urban characteristics. The results reveal coherent and interpretable urban typologies within each city, with some cluster types emerging across both cities despite their differences in scale, density, and cultural context. Comparative analysis showed that adapting the grid size to each city's morphology improves the detection of shared typologies. Simplified clustering based solely on network degree centrality further demonstrated that meaningful structural patterns can be captured even with minimal feature sets. Our findings suggest the presence of functionally convergent urban forms across continents and highlight the importance of spatial scale in cross-city comparisons. The framework offers a scalable and transferable approach for urban analysis, providing valuable insights for planners and policymakers aiming to enhance walkability, accessibility, and well-being. Limitations related to data completeness and feature selection are discussed, and directions for future work—including the integration of additional data sources and human-centered validation—are proposed.

\end{abstract}

\begin{highlights}
\item This study introduces a scalable, open-source clustering framework that leverages OpenStreetMap data to classify and compare neighborhood-level urban forms across international cities.
\item By applying this method to Philadelphia and Lausanne, we identify both shared and city-specific neighborhood typologies shaped by walkability, access to amenities, green space, and multimodal infrastructure.
\item The findings demonstrate the importance of context-aware design and analysis when comparing cities, highlighting how local urban morphology influences the interpretation and transferability of neighborhood cluster patterns.
\end{highlights}




\end{frontmatter}



\section{Introduction}
\label{sec:Introduction}

Urbanization is a global phenomenon that shapes the socioeconomic and environmental landscapes of cities worldwide \cite{haase2018global}. As cities expand and evolve, understanding the underlying patterns of their spatial organization becomes crucial for sustainable urban planning and the enhancement of residents' quality of life \cite{mouratidis2021urban,kent2014three,tavakoli2025psycho,panagopoulos2016urban,barton2009land}. A fundamental aspect of this spatial organization is the concept of urban form, which encompasses the physical layout, infrastructure, land use, and the distribution of amenities within a city \cite{kropf2009aspects,dempsey2010elements,rapoport2013human}. Urban form significantly influences various outcomes, including well-being, walkability, environmental sustainability, and social cohesion \cite{rapoport2013human,frank2001built,jackson2003relationship}.

A pertinent question in urban studies is whether urban forms exhibit similarities across different countries and continents \cite{lemoine2020global,schwanen2002urban,schwarz2010urban,bosselmann2012urban}. Despite the vast diversity in cultural, economic, and geographical contexts, there may exist underlying universal principles that govern the spatial organization of cities. Identifying such similarities can facilitate the transfer of best practices in urban planning and foster a deeper understanding of the factors that contribute to successful urban environments. However, comparing urban forms across diverse contexts poses several challenges, including variations in data availability, differing urban scales, and the complexity of capturing multifaceted urban characteristics \cite{lemoine2020global,schwanen2002urban,schwarz2010urban,bosselmann2012urban}. In this context, there is a pressing need for robust, data-driven methodologies that can objectively evaluate and compare urban forms on a global scale.

This paper presents a novel methodological framework for comparing urban forms across cities enabling comparisons between different countries and continents using advanced clustering algorithms applied on open source available dataset of cities of interest. Leveraging the extensive geospatial data available through OpenStreetMap (OSM), we extract a comprehensive set of features related to urban forms, including topography, multimodality, green spaces, and points of interest. By employing a grid-based approach, cities are then divided into Basic SPatial Units (BSU), each characterized by the extracted features. We then apply Gaussian Mixture Models (GMM) to cluster these BSUs, identifying distinct urban typologies based on their urban form profiles. To demonstrate the efficacy and versatility of our method, we apply it to two geographically and culturally distinct cities: Philadelphia, USA, and Lausanne, Switzerland. These cities differ significantly in size, urban layout, and demographic composition, providing a rigorous test for our clustering approach. The resulting clusters reveal both unique and shared urban form typologies, offering insights into the commonalities and differences in urban forms across continents. The remainder of this paper is structured as follows. Section~\ref{sec:methods} outlines the data collection process, feature extraction, grid generation, and clustering methodology. Section~\ref{sec:results} presents the clustering outcomes for both cities, accompanied by qualitative and quantitative analyses. Finally, Section~\ref{sec:discussion} discusses the implications of our findings, and Section~\ref{sec:conclusion} concludes the paper by summarizing key insights and suggesting directions for future research.

\section{Background and Literature Review}
\label{sec:background}

Understanding urban form is a foundational element in the study of cities, influencing patterns of mobility, land use, accessibility, and quality of life. Urban form broadly encompasses the spatial configuration and structure of built environments, including street networks, land parcels, public spaces, and the distribution of amenities \cite{boeing_measuring_2018,kropf2009aspects,clifton2008quantitative}. The relationship between urban form and walkability, in particular, has been extensively studied due to its implications for sustainability, public health, and transportation planning \cite{barros2017urban,ak2018urban,lee2018planning}. 

It has been found that the structure of the urban form has a significant impact on mobility behavior and the experiences of urban residents. The concept of accessibility is linked primarily to the urban form itself, with key elements including density, mix of land use, and the path network \cite{chapman_smart_2021,ewing_travel_2001,cervero_travel_1997}. The concept of accessibility is often expanded to include additional factors that may impact the experience of the users of the city, including green space, materiality, and urban amenities like benches \cite{fonseca_built_2022}. When taken all together, and when considered from the perspective of the city dweller, it is clear that a diverse and broad range of attributes come together to create the characteristics of urban form. It can therefore be difficult to describe the overall characteristic of a particular part of a city, and even more difficult to compare different places with each other.

Understanding the different types of urban forms that exist within a city can provide critical insights for urban planners and policymakers aiming to enhance walkability, accessibility, and quality of life \cite{hajrasoulih2018urban,barros2017urban,ak2018urban,lee2018planning}. While urban form is a latent construct—meaning it cannot be captured by a single variable but is instead manifested through a combination of measurable indicators such as land use, street networks, and access to amenities—it varies significantly within and across cities \cite{jacobs2015indicators}. In line with this matter, various studies analyzed indicators of urban forms and attempted to classify urban spaces using these indicators. For instance, \cite{wu2024machine} proposed a machine learning-based framework to characterize urban morphology using street patterns, showing that variations in network structure alone can reveal distinct morphological types even in the absence of broader contextual features. In another recent work, \cite{yao2025decoding} analyzed street network data from 326 Chinese cities and identified six core dimensions—such as connectedness, intensity, and network transitivity—that effectively describe variations in urban form. By proposing a streamlined and scalable set of street-based metrics, their framework enables consistent characterization of urban morphology across cities of different sizes. 

However, much of the existing literature has approached this challenge by focusing on a limited set of indicators—such as building density, land use mix, or road network structure—rather than capturing urban form as a holistic and multidimensional construct. This fragmented approach can obscure the complex interactions between built environment features that collectively shape urban experience. As a result, there remains a need for more comprehensive frameworks that incorporate a wide range of spatial, infrastructural, and functional characteristics when classifying urban form through clustering and other data-driven methods. 

Using machine learning and novel artificial intelligence based methods, we can detect urban forms from a set of features in a data-driven fashion. In this regard, clustering approaches emerge as one of the most important methods in identifying urban forms across cities and between different cities. The problem of clustering observations into similar groups without any known labels is a form of unsupervised learning. Clustering applied to spatial data is a particular subset of this problem, and researchers have shown that traditional machine learning algorithms such as k-means, GMMs, and k-medoids can be useful when the data structures allow their use \cite{cao_analyzing_2013, gil_discovery_2012}. More specific methods for spatial data clustering have been developed when data structures are complex or the expected data sizes are prohibitively large \cite{ng_clarans_2002}.

Researchers have found that distinct patterns of urban forms can be found both within and across cities. Analyzing differences in street networks among cities from around the world can reveal insights on regularity of planning as well as which cities are similar to others, though this analysis has been limited to the urban characteristic of the network of paths  \cite{boeing_urban_2019}. Similarly, other researchers have employed clustering to determine similarity of different cities using building footprints \cite{fan_quantifying_2024}. These across-city studies have often looked a large number of cities because their task involves treating each city as an individual observation. Within cities, researchers have taken similar approaches, focusing on single data modalities like street networks or building footprints to cluster neighborhoods \cite{cao_analyzing_2013}. These studies typically focus on neighborhoods within individual cities. However, there remain a need for a clustering methodology that (1) incorporates a wide range of urban form features that would be expected to impact our experiences of being in a city, including our mobility decisions and well-being, and (2) is flexible enough to be able to find clusters of areas both within cities and to compare regions across cities. Here we focus on comparing many regions within two distinct cities in order to demonstrate our method's ability to find clusters within a given city and compare these clusters across cities.

\section{Hypothesis and Research Questions}

This study aims to investigate whether distinct urban form typologies can be systematically identified and compared across geographically and culturally diverse cities using a unified data-driven approach. The motivation stems from a growing interest in understanding the extent to which urban environments—despite being shaped by unique local histories, cultures, and planning practices—may still exhibit common spatial and functional patterns when analyzed through standardized metrics.

\subsection{Research Questions}

The following research questions guide the study:

\begin{enumerate}
    \item \textbf{RQ1:} Can clustering techniques applied to urban form-related features derived from OpenStreetMap effectively identify meaningful neighborhood typologies within individual cities?
    \item \textbf{RQ2:} Are some of these urban typologies shared across different cities, suggesting transferable spatial patterns?
    \item \textbf{RQ3:} How does the choice of spatial resolution (i.e., grid size) influence the identification and comparability of clusters across cities with differing physical, demographic, and infrastructural characteristics?
    \item \textbf{RQ4:} Do similarities in feature distributions across cities contribute to the emergence of common cluster types in geographically distant contexts?
\end{enumerate}

\subsection{Hypotheses}

The study is grounded on the following hypotheses:

\begin{itemize}
    \item \textbf{H1:} GMM clustering applied to spatially distributed urban form features will produce coherent and interpretable clusters within each city, representing distinct neighborhood typologies.
    \item \textbf{H2:} Certain cluster types will be shared across Lausanne and Philadelphia, indicating the presence of globally relevant urban patterns despite differing geographic and cultural contexts.
    \item \textbf{H3:} Allowing for adaptive grid resolution per city will improve the alignment of cluster structures across cities by accounting for variations in urban scale and density, due to the fact that cities spatial distributions are different.
    \item \textbf{H4:} Similarities in urban form-related feature distributions across cities will correlate with the emergence of consistent urban clusters in both cases.
\end{itemize}

\section{Methods}
\label{sec:methods}
The following section outlines our methodology involving data collection, preprocessing, and clustering procedure. We also provide a discussion on the rational behind the choice of number of clusters.

\subsection{Data Collection}
\label{subsec:Data}

The first step in this study is the acquisition of the datasets required for the analysis. The goal is to systematically access a series of features that together can shape urban forms. We utilized \textit{OpenStreetMap}, a freely available and widely used platform that provides comprehensive geospatial data. The primary advantage of using \textit{OpenStreetMap} is its open-access nature, allowing researchers to easily obtain detailed information on urban infrastructure, natural features, and points of interest. Additionally, its global coverage ensures that the methodology developed in this study can be readily adapted to different cities by simply modifying the geographic location, thus enhancing the transferability and scalability of the approach. The code for accessing these features is open source and available at the author's Github repository [citation removed for double blind]. Various features can be captured from \textit{OpenStreetMap} to comprehensively assess various aspects of urban forms. These aspects were categorized into four primary domains, which reflect the multidimensional nature of urban forms. We note that these features are not an exhaustive list and other new features be added to this list; however, for brevity, we only focus on a handful of such features:

\begin{itemize}
    \item \textbf{Topography:} This category encompasses data describing the structure and layout of the street network. Key features include road classifications (e.g., residential, primary, and pedestrian-only streets) and measures of network centrality, which influence the ease and efficiency of pedestrian movement within an urban area.
    \item \textbf{Multi-modality:} Urban forms often interacts with other modes of transportation. This domain captures features such as the number of public transport stops, the presence of cycleways, and crosswalks. These elements illustrate how pedestrian infrastructure integrates with other transportation systems to facilitate seamless mobility.
    \item \textbf{Green Spaces and Natural Elements:} Natural features and greenery enhance walkability by improving environmental quality and providing recreational opportunities. Data in this category include tree density, wooded areas, beaches, and other natural features. Water bodies such as rivers, lakes, and harbors are also included to account for their contribution to the aesthetic and functional aspects of urban spaces.
    \item \textbf{Points of Interest:} This category includes destinations that attract pedestrian activity, such as supermarkets, schools, offices, libraries, and recreational areas. These points serve as critical nodes in origin-destination matrices and influence patterns of pedestrian movement.
\end{itemize}

\subsection{Preprocessing and Generation of Analysis Grid Element}
\label{subsec:grid}
To facilitate spatial clustering, the administrative boundary of the city under analysis is divided into a grid of elementary areas referred to as BSU. In this study, a hexagonal grid structure is employed for defining the BSUs. Hexagonal grids offer distinct advantages, such as uniform distances between neighboring units and the ability to approximate circular regions through expanding rings of neighbors. These properties contribute to smoother spatial paving and reduced variability in outcomes associated with grid size, making hexagonal grids particularly suitable for urban analysis.

In addition to the choice of grid shape, the size of the grid cells is a critical consideration that directly influences the effectiveness of spatial clustering. The grid size must balance two competing objectives: it should be fine enough to capture the local structure and unique urban typologies while remaining large enough to avoid trivial or overly granular classifications. The appropriate grid size varies across cities depending on their specific characteristics and can influence the granularity of the typologies identified. Furthermore, the selection of grid size affects the spatial resolution of the analysis, underscoring the importance of tailoring this parameter to the study context. The determination of grid size can be guided by prior domain knowledge, informed by the characteristics of the urban environment being analyzed. Alternatively, grid size can be optimized based on quantitative metrics that evaluate the quality of clustering, such as silhouette scores or other measures of spatial cohesion and separation \cite{shahapure2020cluster}. In this study, to determine the optimal grid size, we utilize the silhouette score \cite{shahapure2020cluster}, a metric that assesses the quality of clustering by measuring the cohesion within clusters and the separation between them. The silhouette score is defined as:

\begin{equation} \text{Silhouette Score} = \frac{b - a}{\max(a, b)} \label{eq:silhouette} \end{equation}

where: \begin{itemize} \item $a$ is the average distance between a data point and all other points in the same cluster, \item $b$ is the average distance between a data point and all points in the nearest neighboring cluster. \end{itemize}

A higher silhouette score indicates that the data points are well-clustered, with clear boundaries between clusters. By maximizing the silhouette score, we select the grid size that yields the most coherent and well-separated clusters, ensuring that the discretization size is both meaningful and effective for the analysis. 

We note that our primary focus is not on determining the optimal size of the BSU, but rather on demonstrating the flexibility and applicability of the clustering methodology itself. However, it is important to acknowledge that the choice of BSU size can influence the granularity of the resulting clusters, as finer grids capture more localized urban features while coarser grids aggregate broader spatial patterns. The presented framework is adaptable to any BSU size, and users can tailor the grid resolution depending on the specific objectives of their study, the scale of the city, or the level of detail desired in the analysis

For each BSU, we calculate values for the selected features based on the data available within its boundaries. These values can be aggregated as total counts for extensive features (e.g., the number of trees or public transport stops) or as average values for intensive features (e.g., road centrality measures). This ensures that the data for each feature is appropriately represented and standardized across all BSUs. Once the data has been processed and aggregated, it is organized into a matrix format, where rows correspond to individual BSUs and columns represent the selected urban form related features. 

\subsection{Clustering}
In order to perform clustering, we leverage GMM due to their efficacy in modeling complex, multimodal distributions commonly present in urban spatial data. GMMs are applied to BSUs based on their aggregated urban form features. GMMs are well-known for their flexibility in representing data as a combination of multiple Gaussian distributions, each corresponding to a distinct cluster or underlying generative process.

Mathematically, the probability density function of a GMM is expressed as:

\begin{equation} p(\mathbf{x}|\lambda) = \sum_{k=1}^{K} \pi_k \mathcal{N}(\mathbf{x}|\boldsymbol{\mu}_k, \boldsymbol{\Sigma}_k), \label{eq:gmm} \end{equation}

where: \begin{itemize} \item $\mathbf{x}$ is a data point in the feature space, \item $K$ is the number of Gaussian components, \item $\pi_k$ represents the mixing coefficient for the $k$-th Gaussian component, satisfying $\sum_{k=1}^{K} \pi_k = 1$, \item $\mathcal{N}(\mathbf{x}|\boldsymbol{\mu}_k, \boldsymbol{\Sigma}_k)$ denotes the multivariate Gaussian distribution with mean vector $\boldsymbol{\mu}_k$ and covariance matrix $\boldsymbol{\Sigma}_k$. \end{itemize}

The Expectation-Maximization (EM) algorithm is utilized to estimate the parameters $\lambda = {\pi_k, \boldsymbol{\mu}_k, \boldsymbol{\Sigma}k}{k=1}^{K}$ of the mixture model. The EM algorithm iteratively performs two main steps: \begin{enumerate} \item \textbf{Expectation (E-step)}: Calculates the posterior probabilities (responsibilities) that each BSU belongs to each Gaussian component. \item \textbf{Maximization (M-step)}: Updates the parameters $\lambda$ to maximize the expected log-likelihood of the data given the current responsibilities. \end{enumerate}

The iterative process continues until convergence, resulting in parameter estimates that best fit the observed feature data.

\subsection{Determining the Number of Clusters}
\label{subsec:clusters} Reflective of underlying urban forms, the number of clusters, corresponding to the number of Gaussian distributions in the mixture model, is a fundamental hyperparameter in GMMs. To identify the optimal number of clusters, we employ the Bayesian Information Criterion (BIC), which provides a balance between model fit and complexity \cite{zhao2015mixture}. The BIC is mathematically defined as:

\begin{equation} BIC = k \ln(N) - 2 \ln(\hat{\mathcal{L}}), \label{eq:BIC} \end{equation}

where: \begin{itemize} \item $k$ is the number of parameters in the model, \item $N$ is the number of data points, \item $\hat{\mathcal{L}}$ denotes the maximized value of the likelihood function for the model. \end{itemize}. 

By calculating the BIC for various numbers of clusters, we select the model that minimizes the BIC value. This selection criterion ensures an optimal balance between the accuracy of the clustering and the simplicity of the model, thereby mitigating the risk of overfitting by favoring models with fewer clusters when appropriate.

\section{Case Study of Philadelphia and Lausanne}
\label{sec:results}

Based on the developed framework, we are interested in applying it to two cities located on different continents, each with distinct urban forms, population densities, and cultural influences. Lausanne, situated in Switzerland, represents a compact, European city characterized by its hilly terrain, historical architecture, and pedestrian-friendly urban planning. In contrast, Philadelphia, located in the United States, is a larger metropolitan area with a grid-based urban layout, diverse neighborhoods, and a mix of historic and modern infrastructure. These differences provide an ideal opportunity to test the robustness of the framework in capturing urban typologies and comparing urban form features across geographically and culturally distinct contexts. By applying the developed clustering approach, we aim to identify patterns that reveal the extent to which urban forms characteristics align or differ between Lausanne and Philadelphia. This comparison not only highlights how cities on different continents meet the needs of their inhabitants but also informs whether universal urban design principles can be derived from these observations.

We start by retrieving the data features based on the framework provided in the methodology. More specifically, the selected features for each category are summarized in Table 1, offering a comprehensive overview of the variables used to assess urban forms in the context of this research. These features are extracted for Lausanne and Philadelphia.

\resizebox{0.98\textwidth}{!}{
\begin{tabular}{|p{2.5cm}|p{3.5cm}|p{14cm}|}
\hline
\textbf{Category}       & \textbf{Subcategory} & \textbf{Feature} \\ \hline

\multicolumn{3}{|c|}{\textbf{Topography}} \\ \hline
Highway & Pedestrian    & \texttt{city.count\_feature(['highway','pedestrian'])} \\ \cline{2-3}
        & Service       & \texttt{city.count\_feature(['highway','service'])} \\ \cline{2-3}
        & Living Street & \texttt{city.count\_feature(['highway','living\_street'])} \\ \cline{2-3}
        & Footway       & \texttt{city.count\_feature(['highway','footway'])} \\ \cline{2-3}
        & Steps         & \texttt{city.count\_feature(['highway','steps'])} \\ \cline{2-3}
        & Path          & \texttt{city.count\_feature(['highway','path'])} \\ \hline
Walk    & Degree Centrality & \texttt{city.network\_feature('walk', 'degree\_centrality','extensive')} \\ \hline

\multicolumn{3}{|c|}{\textbf{Multimodality}} \\ \hline
Public Transport & -             & \texttt{city.count\_feature(['public\_transport'])} \\ \hline
Highway          & Bus Stop      & \texttt{city.count\_feature(['highway','bus\_stop'])} \\ \cline{2-3}
                 & Cycleway      & \texttt{city.count\_feature(['highway','cycleway'])} \\ \cline{2-3}
                 & Crossing      & \texttt{city.count\_feature(['highway','crossing'])} \\ \hline
Railway          & Subway Entrance & \texttt{city.count\_feature(['railway','subway\_entrance'])} \\ \hline

\multicolumn{3}{|c|}{\textbf{Points of Interest}} \\ \hline
Building & Residential   & \texttt{city.count\_feature(['building','residential'])} \\ \cline{2-3}
         & Commercial    & \texttt{city.count\_feature(['building','commercial'])} \\ \cline{2-3}
         & Public        & \texttt{city.count\_feature(['building','public'])} \\ \cline{2-3}
         & School        & \texttt{city.count\_feature(['building','school'])} \\ \cline{2-3}
         & Church        & \texttt{city.count\_feature(['building','church'])} \\ \cline{2-3}
         & University    & \texttt{city.count\_feature(['building','university'])} \\ \cline{2-3}
         & Train Station & \texttt{city.count\_feature(['building','train\_station'])} \\ \hline
Amenity  & Parking       & \texttt{city.count\_feature(['amenity','parking'])} \\ \cline{2-3}
         & Restaurant    & \texttt{city.count\_feature(['amenity','restaurant'])} \\ \cline{2-3}
         & Cafe          & \texttt{city.count\_feature(['amenity','cafe'])} \\ \cline{2-3}
         & Bar           & \texttt{city.count\_feature(['amenity','bar'])} \\ \cline{2-3}
         & Pub           & \texttt{city.count\_feature(['amenity','pub'])} \\ \cline{2-3}
         & Theatre       & \texttt{city.count\_feature(['amenity','theatre'])} \\ \cline{2-3}
         & Cinema        & \texttt{city.count\_feature(['amenity','cinema'])} \\ \cline{2-3}
         & Library       & \texttt{city.count\_feature(['amenity','library'])} \\ \cline{2-3}
         & Hospital      & \texttt{city.count\_feature(['amenity','hospital'])} \\ \cline{2-3}
         & Pharmacy      & \texttt{city.count\_feature(['amenity','pharmacy'])} \\ \cline{2-3}
         & Doctors       & \texttt{city.count\_feature(['amenity','doctors'])} \\ \hline

\multicolumn{3}{|c|}{\textbf{Natural elements}} \\ \hline
Natural & -   & \texttt{city.count\_feature(['natural'])} \\ \hline

\end{tabular}
}

Based on the silhouette score analysis proposed in our framework, the following grid sizes were chosen for this study. :\\

\begin{table}[h!]
\centering
\begin{tabular}{l r}
\toprule
\textbf{City} & \textbf{Grid size (m)} \\
\midrule
Philadelphia & 1500 \\
Lausanne     & 450  \\
\bottomrule
\end{tabular}
\caption{Grid size of selected cities.}
\label{tab:size_grid}
\end{table}

\begin{figure}[H]
    \begin{subfigure}{.5\textwidth}
    \includegraphics[width=7cm]{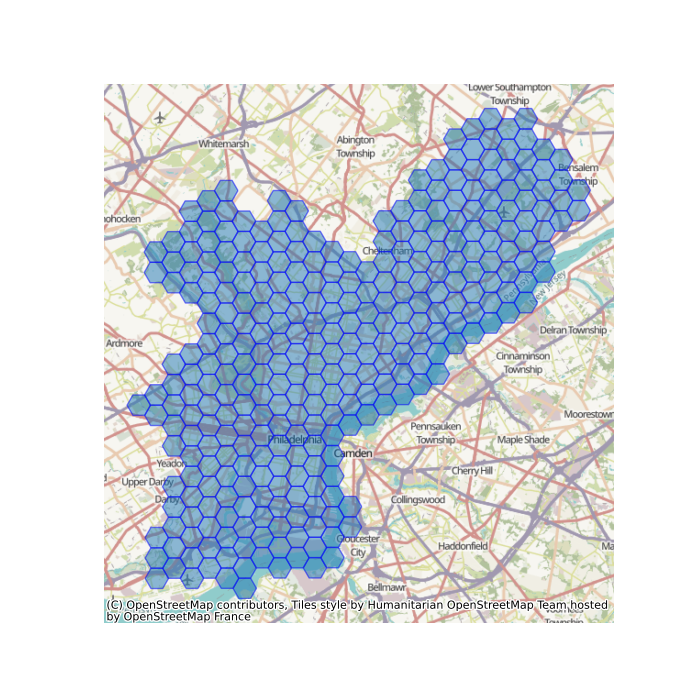}
    \caption{}\label{fig:grid_philadelphia}
    \end{subfigure}
    \begin{subfigure}{.5\textwidth}
    \includegraphics[width=7cm]{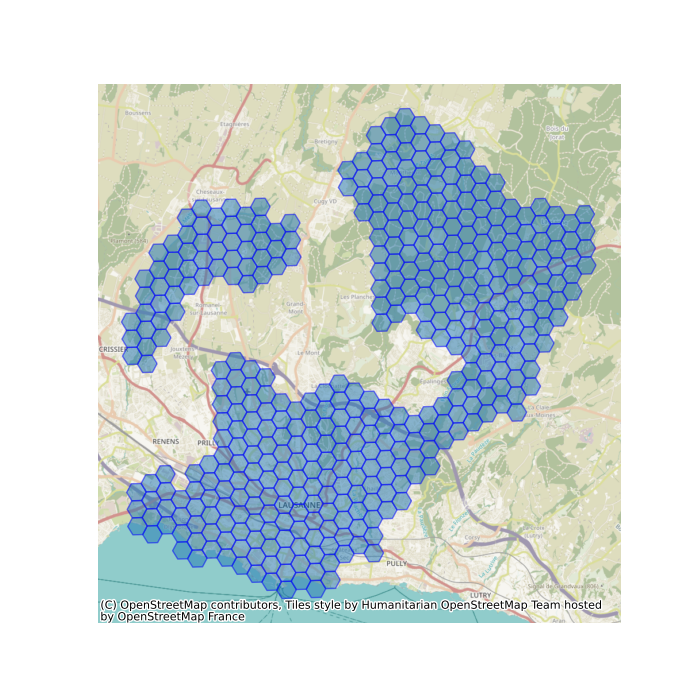}
    \caption{}\label{fig:grid_lausanne}
    \end{subfigure}
    \caption{Grid of hexagonal BSU for (a) Philadelphia with grid size of 1500 m and (b) Lausanne with grid size 450 m.}
\end{figure}

Following our analysis framework, the optimal number of cluster for the selected grid size for both cities is given in the table \ref{tab:size_grid_clusters}. 

\begin{table}[h!]
\centering
\begin{tabular}{l r r}
\toprule
\textbf{City} & \textbf{Grid size (m)} & \textbf{\# clusters (-)}\\
\midrule
Philadelphia & 1500 & 7\\
Lausanne     & 450 & 5\\
\bottomrule
\end{tabular}
\caption{Grid size of selected cities.}
\label{tab:size_grid_clusters}
\end{table}

\begin{figure}[H]
    \begin{subfigure}{.5\textwidth}
        \centering
        \includegraphics[width=7cm]{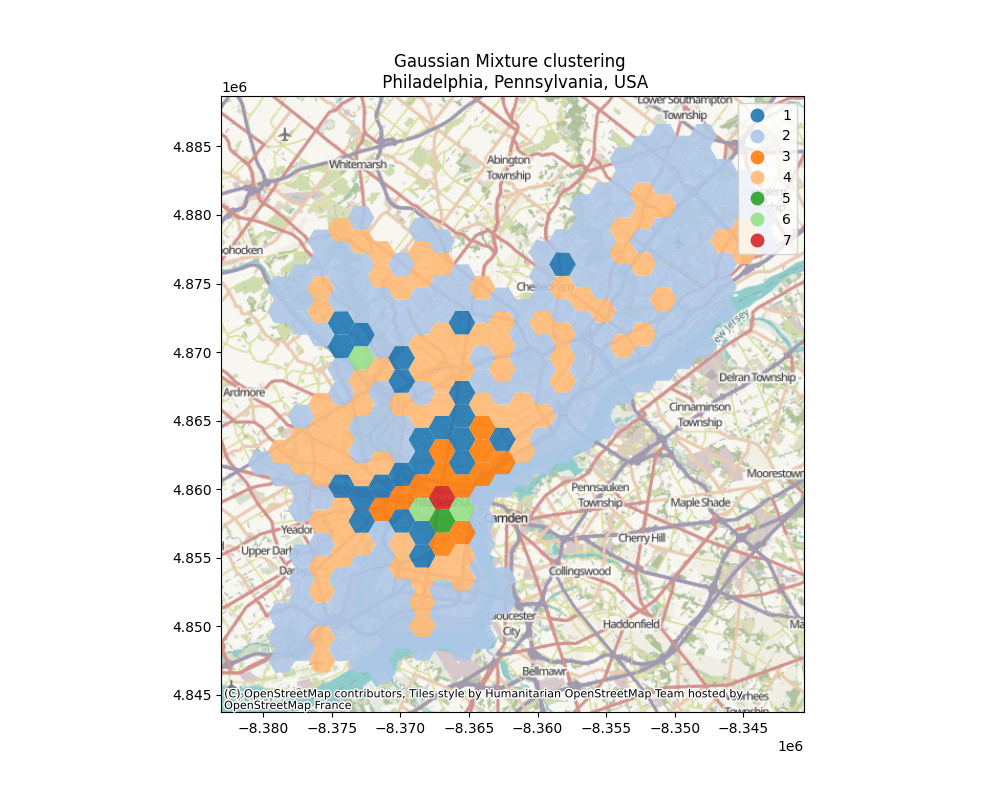}
        \caption{}
        \label{fig:gmm_philadelphia_7}
    \end{subfigure}
    \begin{subfigure}{.5\textwidth}
        \centering
        \includegraphics[width=7cm]{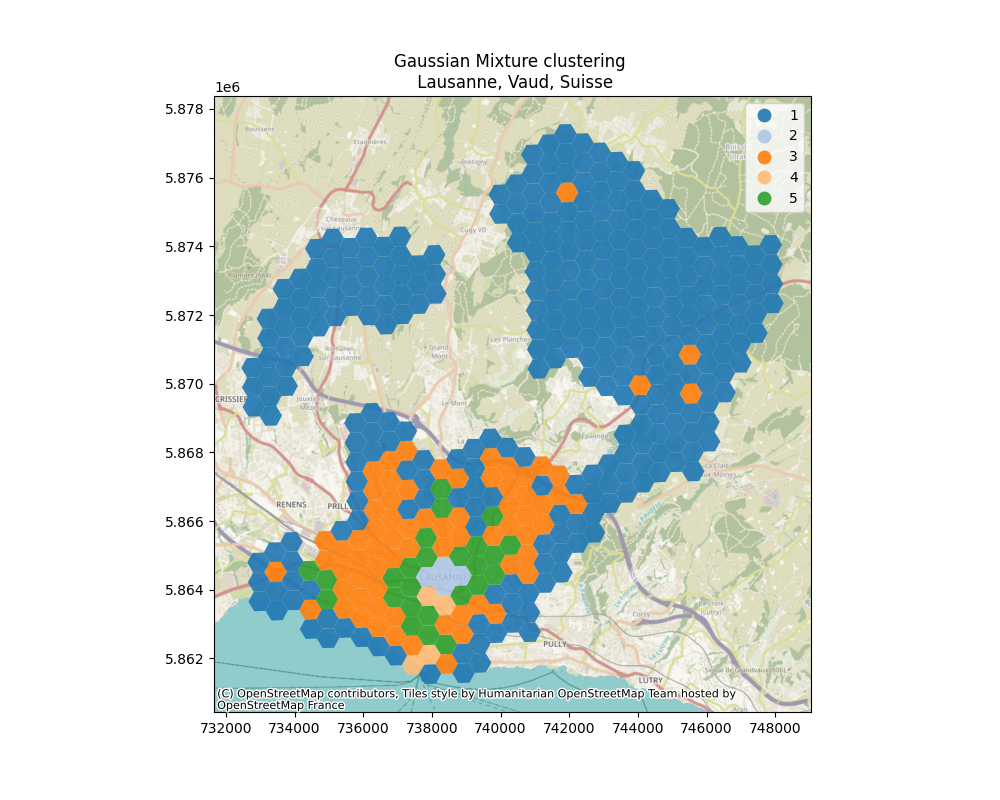}
        \caption{}
        \label{fig:gmm_lausanne_5}
    \end{subfigure}
    \caption{Results of GMM clustering for (a) Philadelphia with 8 clusters (grid size of 1500\,m) and (b) Lausanne with 5 clusters (grid size of 450\,m).}
    \label{fig:clustering_examples}
\end{figure}

\subsection{Quantitative Analysis of Cluster Composition}
\label{subsec:quantitative_clusters}

\begin{figure}[H]
    \centering
    \includegraphics[width=0.95\textwidth]{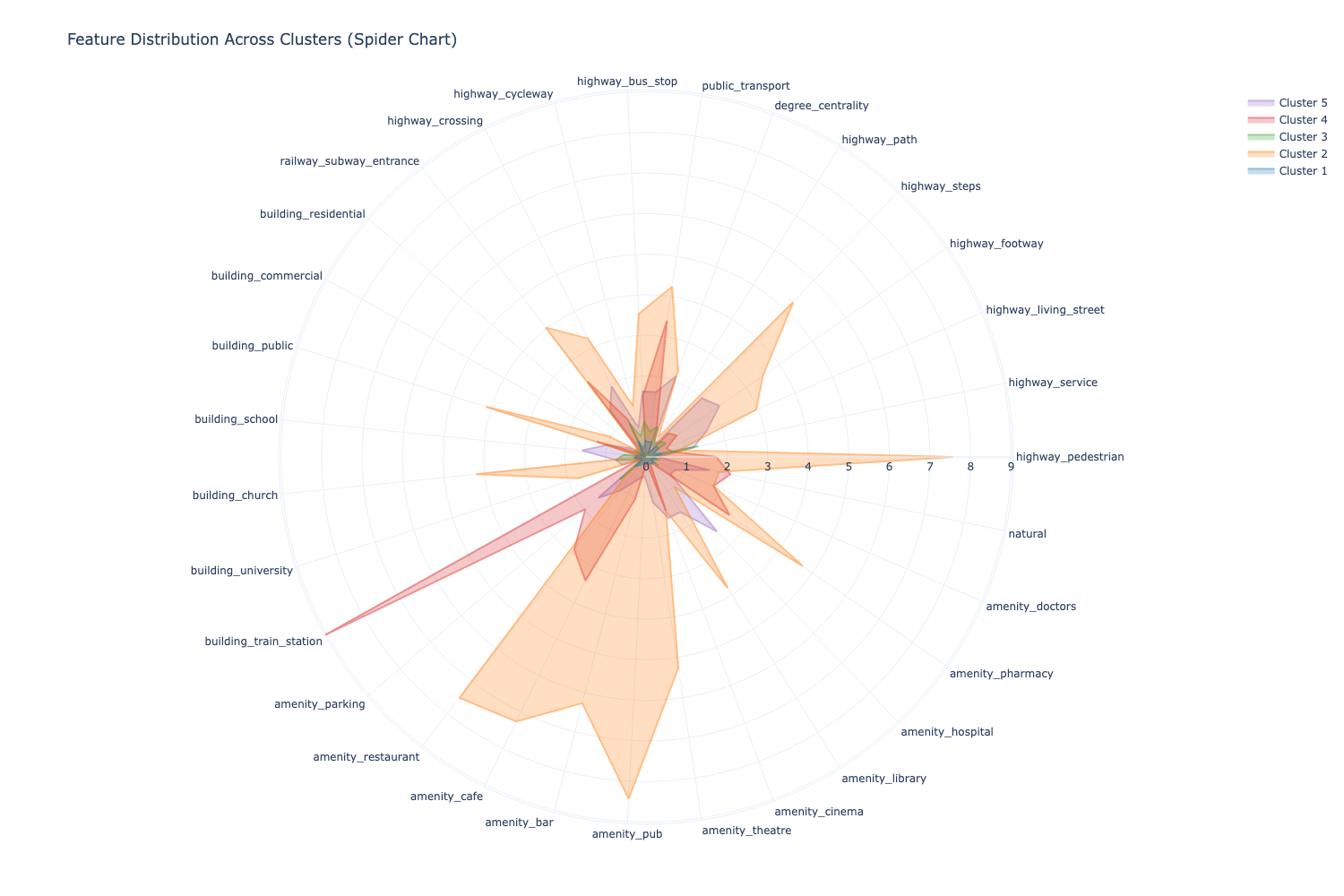}
    \caption{Spider diagram for the GMM with 5 clusters (Lausanne).}
    \label{fig:spider_gmm_lausanne}
\end{figure}

\begin{figure}[H]
    \centering
    \includegraphics[width=0.95\textwidth]{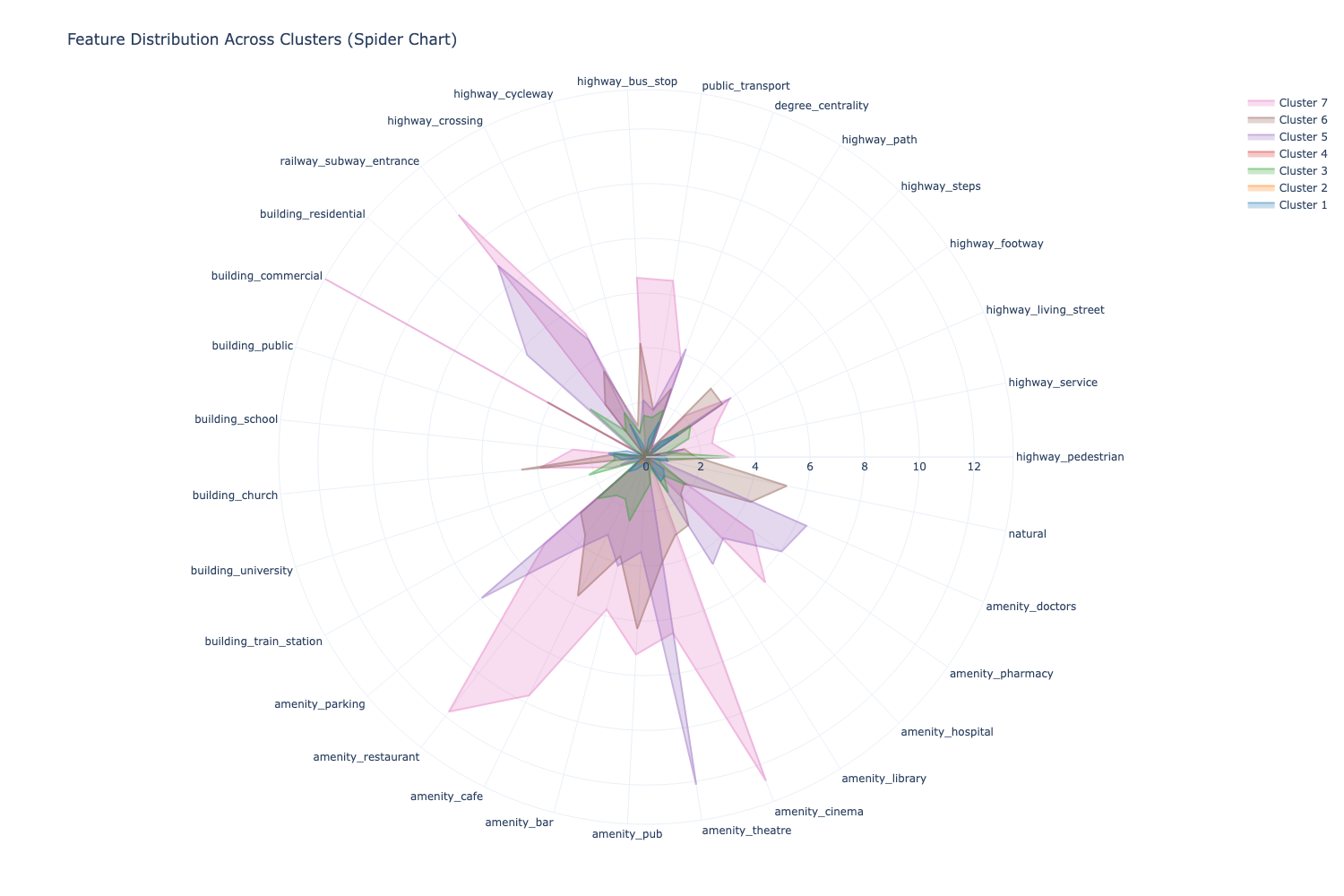}
    \caption{Spider diagram for the GMM with 8 clusters (Philadelphia).}
    \label{fig:spider_gmm_phily}
\end{figure}

To further explore the nature of the identified clusters, we employ histogram representations of the feature distributions. Figures~\ref{fig:spider_gmm_lausanne} and \ref{fig:spider_gmm_phily} present stacked histograms depicting the relative importance of the various urban form-related features within each cluster. The normalized feature values signify each variable's contribution to assigning a BSU to a specific cluster. These visual representations facilitate a comparative analysis of cluster characteristics, highlighting which features exert the strongest influence on the clustering outcomes and provides a more comprehensive understanding of the underlying patterns distinguishing one cluster from another. A detailed interpretation of the clustering results for both cities is given next.

\paragraph{\textbf{Lausanne}} For Lausanne, the most immediately noticeable observation from Fig. \ref{fig:spider_gmm_lausanne} is that, although five clusters were identified, only four are clearly visible on the graph. Cluster 1 (blue) is nearly absent. This can be explained by referring to Table \ref{tab:lausanne_data_cluster}, which shows that Cluster 1 is characterized by uniformly low values across all features. This pattern is consistent with the spatial distribution in Fig. \ref{fig:gmm_lausanne_5}, where Cluster 1 corresponds to largely unoccupied or undeveloped areas with minimal urban activity.

Cluster 2 exhibits the highest values across several key features, particularly in the food and beverage category (e.g., restaurants, cafés, bars, and pubs). It is also characterized by a high density of pedestrian streets, public buildings, and public transportation stops. As shown in Fig. \ref{fig:gmm_lausanne_5}, this cluster corresponds to the Hypercenter of Lausanne, a highly active urban area. The grouping of these features into a single cluster aligns with expectations, indicating that the clustering algorithm effectively captured the spatial correlation between the concentration of bars, pedestrian-friendly infrastructure (e.g., terraces), and accessibility via public transportation.

Cluster 4 is particularly noteworthy due to its high values near the main train station, including parking facilities, several restaurants, and a moderate presence of natural features. As shown in Fig. \ref{fig:gmm_lausanne_5}, this cluster corresponds to the area surrounding Lausanne’s central train station and the adjacent boat terminal. The presence of natural elements can be attributed to the boat station’s proximity to the lake and surrounding vegetation. Notably, this cluster also includes a significant number of restaurants, suggesting a spatial association between major transportation hubs and food service establishments.

Cluster 3 is characterized by generally lower values across most features, with modest presence of public transportation stops, churches, and schools. As seen in Fig. \ref{fig:gmm_lausanne_5}, this cluster represents one of the more typical residential neighborhoods outside the city center and comprises the more dispersed areas of Lausanne. Its composition reflects a lower-density urban fabric, with fewer commercial and transportation-related amenities compared to central clusters.

Finally, Cluster 5, which forms the periphery of the hypercenter, is distinguished by a high density of points of interest, including cinemas, theaters, libraries, hospitals, and active local streets. It is also characterized by a significant number of bus stops and pedestrian crossings. This spatial pattern aligns with expectations for areas surrounding the city center, where a diverse range of urban activities is concentrated. The distribution of public transport infrastructure in this cluster further supports its role as a transitional zone connecting the hypercenter with surrounding neighborhoods.

\begin{table}[htbp]
\centering
\resizebox{\textwidth}{!}{%
\renewcommand{\arraystretch}{1.2}
\setlength{\tabcolsep}{6pt}
\begin{tabular}{lccccc}
\toprule
\textbf{Feature} & \textbf{Cluster 1} & \textbf{Cluster 2} & \textbf{Cluster 3} & \textbf{Cluster 4} & \textbf{Cluster 5} \\
\midrule
highway\_pedestrian & 0.170 & 7.569 & 0.117 & 1.731 & 0.286 \\
highway\_service & 0.422 & 0.813 & 1.316 & 0.662 & 1.200 \\
highway\_living\_street & 0.194 & 2.952 & 0.078 & 0.546 & 1.622 \\
highway\_footway & 0.373 & 3.511 & 0.596 & 0.919 & 2.200 \\
highway\_steps & 0.358 & 5.262 & 0.510 & 0.806 & 1.995 \\
highway\_path & 0.104 & 0.567 & 0.388 & 0.072 & 0.277 \\
degree\_centrality & 0.394 & 2.274 & 0.794 & 0.754 & 2.118 \\
public\_transport & 0.375 & 4.246 & 0.629 & 3.397 & 1.613 \\
highway\_bus\_stop & 0.398 & 3.530 & 0.873 & 1.503 & 1.617 \\
highway\_cycleway & 0.198 & 1.292 & 0.524 & 0.012 & 0.739 \\
highway\_crossing & 0.426 & 3.252 & 0.923 & 1.032 & 1.936 \\
railway\_subway\_entrance & 0.174 & 4.033 & 0.104 & 2.350 & 1.441 \\
building\_residential & 0.070 & 0.198 & 0.296 & 0.194 & 0.011 \\
building\_commercial & 0.117 & 1.052 & 0.309 & 0.164 & 0.420 \\
building\_public & 0.151 & 4.134 & 0.007 & 1.277 & 0.991 \\
building\_school & 0.255 & 0.077 & 0.545 & 0.116 & 1.583 \\
building\_church & 0.285 & 4.210 & 0.750 & 0.285 & 0.743 \\
building\_university & 0.126 & 1.748 & 0.238 & 0.126 & 0.623 \\
building\_train\_station & 0.081 & 0.110 & 0.110 & 9.055 & 0.110 \\
amenity\_parking & 0.356 & 0.549 & 0.855 & 1.970 & 1.545 \\
amenity\_restaurant & 0.245 & 7.510 & 0.112 & 2.890 & 1.053 \\
amenity\_cafe & 0.200 & 7.264 & 0.031 & 3.394 & 0.685 \\
amenity\_bar & 0.124 & 6.268 & 0.062 & 1.075 & 0.537 \\
amenity\_pub & 0.170 & 8.433 & 0.081 & 0.240 & 0.486 \\
amenity\_theatre & 0.152 & 5.257 & 0.019 & 0.170 & 1.132 \\
amenity\_cinema & 0.134 & 1.415 & 0.134 & 1.415 & 1.601 \\
amenity\_library & 0.201 & 3.807 & 0.111 & 0.201 & 1.595 \\
amenity\_hospital & 0.213 & 1.014 & 0.008 & 0.213 & 2.536 \\
amenity\_pharmacy & 0.238 & 4.701 & 0.372 & 2.506 & 0.728 \\
amenity\_doctors & 0.131 & 1.810 & 0.085 & 1.810 & 0.801 \\
natural & 0.261 & 1.834 & 0.346 & 2.121 & 1.602 \\
\bottomrule
\end{tabular}%
}
\caption{Values per cluster for Lausanne}
\label{tab:lausanne_data_cluster}
\end{table}

\paragraph{\textbf{Philadelphia}} In the case of Philadelphia, one immediately noticeable observation is that, although seven clusters were identified, only five are visibly represented on the graph. Clusters 2 (orange) and 4 (red) are nearly absent. This observation is supported by Table \ref{tab:philadelphia_data_cluster}, which shows that these two clusters exhibit significantly lower values across all features compared to the others. Cluster 4, as seen in Fig. \ref{fig:gmm_philadelphia_7}, corresponds to areas dominated by major highways—an element not included in the clustering features—explaining its lack of prominence on the graph. Cluster 2, on the other hand, represents the city’s peripheral residential areas, which are characterized by low point-of-interest density and limited transportation infrastructure.

The most immediately striking cluster is Cluster 7, which stands out due to its relatively high concentration of restaurants, bars, pubs, cinemas, and theaters—indicators of dynamic and culturally active areas. It also exhibits a high density of commercial buildings and extensive public transportation infrastructure, including subway stations. As shown in Fig. \ref{fig:gmm_philadelphia_7}, this cluster consists of a single cell, suggesting that it corresponds almost exactly to the hypercenter of Philadelphia, where commercial, cultural, and transit activities are highly concentrated.

Another cluster exhibiting notable feature values is Cluster 5. This cluster is characterized by a high density of theaters, parking facilities, doctors’ offices, and pharmacies. It also includes a significant number of residential buildings and subway entrances, suggesting a mixed-use urban environment. An interesting dynamic emerges from this composition: the strong presence of parking in conjunction with medical services implies the need to accommodate patients who may have limited access to or difficulty using public transportation. Conversely, the presence of theaters and subway entrances points to a more transit-oriented and culturally active context. As shown in Fig. \ref{fig:gmm_philadelphia_7}, this cluster also consists of a single cell, located adjacent to the city’s hypercenter.

Cluster 6 is characterized by a high concentration of pubs, cafés, bus stops, and natural elements. This composition aligns with areas where individuals are more likely to rely on public transportation to access leisure and outdoor spaces. An interesting observation from Fig. \ref{fig:gmm_philadelphia_7} is that this cluster comprises three spatially non-contiguous cells, demonstrating the algorithm’s ability to identify similar urban characteristics across non-adjacent areas. The relatively high number of amenities within this cluster can be partially attributed to the fact that two of the three cells are located near the city center, where amenity density is typically higher.

Cluster 3 exhibits generally lower values compared to the previously discussed clusters, but it maintains moderate representation across a broad range of features. As illustrated in Fig. \ref{fig:gmm_philadelphia_7}, the cells within this cluster correspond primarily to residential areas, characterized by the presence of universities, parking facilities, and a notable number of pedestrian streets. The distribution of features suggests that the clustering algorithm effectively captures the typical characteristics of residential neighborhoods, including educational institutions and housing infrastructure.

\begin{table}[htbp]
\centering
\resizebox{\textwidth}{!}{%
\renewcommand{\arraystretch}{1.2}
\setlength{\tabcolsep}{6pt}
\begin{tabular}{lccccccc}
\toprule
\textbf{Feature} & \textbf{Cluster 1} & \textbf{Cluster 2} & \textbf{Cluster 3} & \textbf{Cluster 4} & \textbf{Cluster 5} & \textbf{Cluster 6} & \textbf{Cluster 7} \\
\midrule
highway\_pedestrian & 0.720 & 0.260 & 3.035 & 0.163 & 0.530 & 1.805 & 3.263 \\
highway\_service & 0.341 & 0.363 & 0.788 & 0.521 & 1.405 & 1.432 & 2.465 \\
highway\_living\_street & 0.036 & 0.087 & 1.687 & 0.087 & 0.087 & 0.087 & 2.752 \\
highway\_footway & 1.449 & 0.383 & 1.980 & 0.034 & 3.800 & 3.398 & 3.715 \\
highway\_steps & 0.774 & 0.270 & 0.690 & 0.144 & 0.379 & 3.441 & 2.088 \\
highway\_path & 0.017 & 0.054 & 0.012 & 0.096 & 0.366 & 0.343 & 0.299 \\
degree\_centrality & 1.366 & 0.479 & 1.826 & 0.255 & 4.214 & 2.666 & 3.683 \\
public\_transport & 0.652 & 0.370 & 1.459 & 0.287 & 1.719 & 1.745 & 6.514 \\
highway\_bus\_stop & 0.136 & 0.280 & 1.510 & 0.124 & 2.082 & 4.170 & 6.557 \\
highway\_cycleway & 0.466 & 0.039 & 0.920 & 0.229 & 0.863 & 1.168 & 0.051 \\
highway\_crossing & 1.347 & 0.375 & 1.801 & 0.028 & 4.759 & 3.494 & 5.014 \\
railway\_subway\_entrance & 0.153 & 0.189 & 1.186 & 0.117 & 8.845 & 2.430 & 11.202 \\
building\_residential & 0.076 & 0.192 & 2.692 & 0.002 & 5.720 & 0.031 & 0.025 \\
building\_commercial & 0.046 & 0.139 & 0.306 & 0.060 & 0.491 & 4.124 & 13.435 \\
building\_public & 0.710 & 0.057 & 0.057 & 0.057 & 0.057 & 0.057 & 0.057 \\
building\_school & 1.382 & 0.285 & 1.169 & 0.010 & 1.080 & 1.080 & 2.684 \\
building\_church & 0.885 & 0.230 & 1.156 & 0.110 & 0.400 & 4.583 & 3.886 \\
building\_university & 0.080 & 0.136 & 2.181 & 0.059 & 0.975 & 0.048 & 1.254 \\
building\_train\_station & 0.367 & 0.150 & 0.344 & 0.202 & 0.184 & 0.184 & 0.184 \\
amenity\_parking & 0.826 & 0.378 & 2.332 & 0.023 & 7.917 & 3.144 & 4.887 \\
amenity\_restaurant & 0.560 & 0.279 & 1.769 & 0.107 & 4.329 & 3.631 & 11.775 \\
amenity\_cafe & 0.372 & 0.245 & 1.709 & 0.158 & 3.164 & 5.657 & 9.708 \\
amenity\_bar & 0.259 & 0.267 & 2.410 & 0.070 & 4.104 & 3.738 & 5.752 \\
amenity\_pub & 0.418 & 0.275 & 1.366 & 0.047 & 3.477 & 6.291 & 7.229 \\
amenity\_theatre & 0.076 & 0.181 & 0.998 & 0.121 & 12.121 & 3.896 & 6.513 \\
amenity\_cinema & 0.134 & 0.134 & 0.398 & 0.022 & 0.134 & 3.057 & 12.628 \\
amenity\_library & 1.030 & 0.435 & 1.530 & 0.325 & 4.617 & 2.933 & 0.435 \\
amenity\_hospital & 0.979 & 0.240 & 0.923 & 0.055 & 4.092 & 1.855 & 6.329 \\
amenity\_pharmacy & 0.768 & 0.313 & 1.804 & 0.045 & 6.044 & 1.695 & 4.739 \\
amenity\_doctors & 0.032 & 0.222 & 0.573 & 0.188 & 6.381 & 4.168 & 0.257 \\
natural & 0.826 & 0.167 & 0.496 & 0.116 & 0.335 & 5.255 & 0.639 \\
\bottomrule
\end{tabular}%
}
\caption{Values per cluster for Philadelphia}
\label{tab:philadelphia_data_cluster}
\end{table}

\subsection{Simple dataset, network centrality}\label{appendix:simple_dataset}

We also explored a simplified version of our clustering analysis by considering only a single key metric—network degree centrality—which captures the connectivity density of the street network. This exploration serves two purposes: first, to evaluate whether meaningful urban typologies can still be identified using minimal feature sets; and second, to understand the role of street network structure alone in shaping urban form. Degree centrality was selected because it is a dense, globally relevant measure that directly reflects urban permeability, accessibility, and connectivity.

\begin{figure}[H]
    \begin{subfigure}{.5\textwidth}
    \includegraphics[width=7cm]{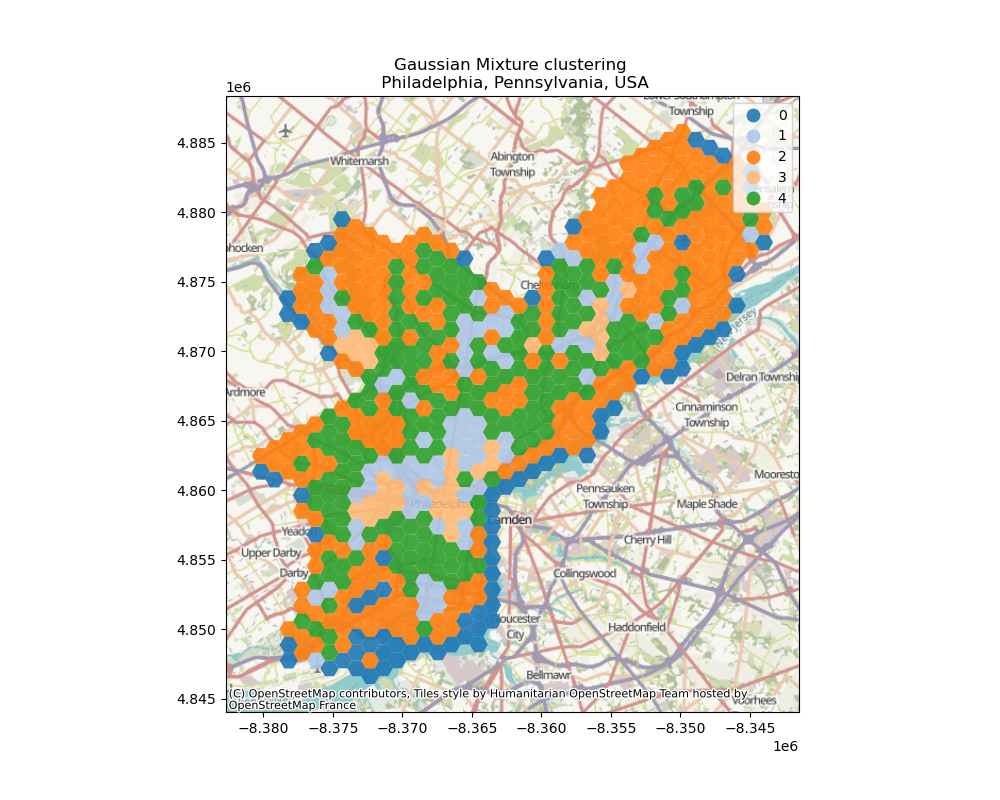}
    \caption{}\label{fig:network_clusters}
    \end{subfigure}
    \begin{subfigure}{.5\textwidth}
    \includegraphics[width=7cm]{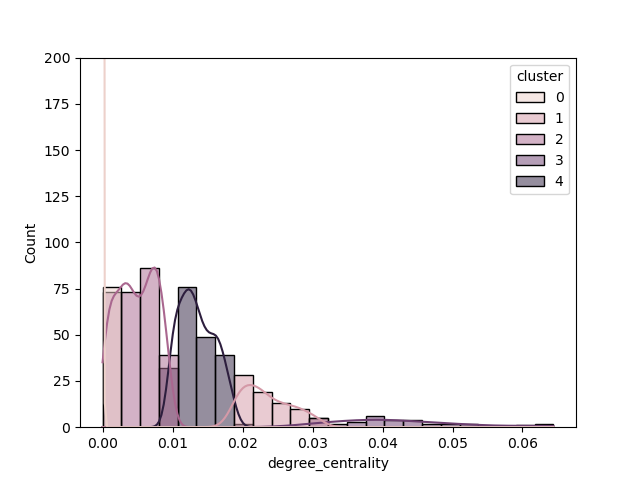}
    \caption{}\label{fig:network_distrib}
    \end{subfigure}
    \caption{(a) Visualization of the result of clustering. (b) Distribution of the feature across clusters.}
\end{figure}

When clustering BSUs based solely on degree centrality, the method produced results that were highly interpretable and strongly aligned with known spatial structures. As illustrated in Figure~\ref{fig:network_clusters}, the clusters generated for the city of Philadelphia corresponded clearly to variations in street network density, distinguishing between highly connected urban cores and more fragmented or suburban peripheral areas.

Furthermore, the distribution of degree centrality values across the identified clusters exhibited a less skewed and more even profile compared to multi-feature clustering (Figure~\ref{fig:network_distrib}). This smoother distribution suggests that degree centrality alone provides a strong, consistent signal for differentiating urban spatial structures without the added complexity or noise introduced by multiple heterogeneous features.

Overall, these results demonstrate that even when constrained to a single feature, the proposed framework remains effective for uncovering meaningful urban form typologies. While a richer feature set allows for more nuanced interpretations, the degree centrality-only analysis highlights the fundamental role of street network structure as a backbone of urban morphology.

\subsection{Qualitative Validation of Clusters}
\label{subsec:qualitative_clusters}

A qualitative approach to evaluate the validity of the identified clusters involves comparing satellite imagery of randomly selected cells within the same cluster. If these spatially distant cells exhibit similar structural or land-use characteristics, it suggests that the clustering algorithm is effectively capturing meaningful and consistent urban features. In the following examples, the selected cells are intentionally located far apart within each city to reduce the influence of spatial proximity and better isolate the role of shared feature patterns in driving the clustering outcomes.

\begin{figure}[H]
    \centering
    \begin{subfigure}[b]{0.48\linewidth}
        \centering
        \includegraphics[width=\textwidth]{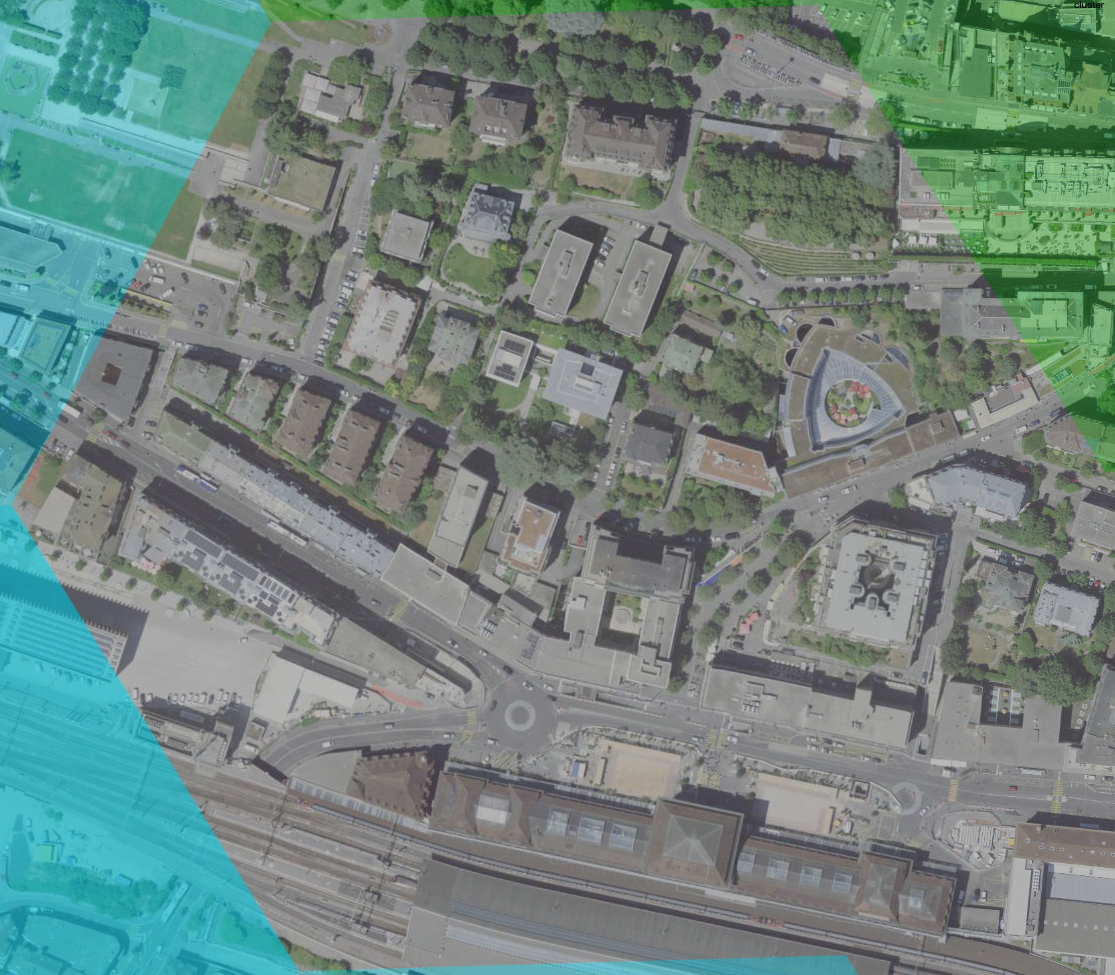}
        \caption{Satellite view of a first cell in Cluster~4 (Lausanne).}
        \label{fig:sat_view_1_lausanne}
    \end{subfigure}
    \hfill
    \begin{subfigure}[b]{0.48\linewidth}
        \centering
        \includegraphics[width=\textwidth]{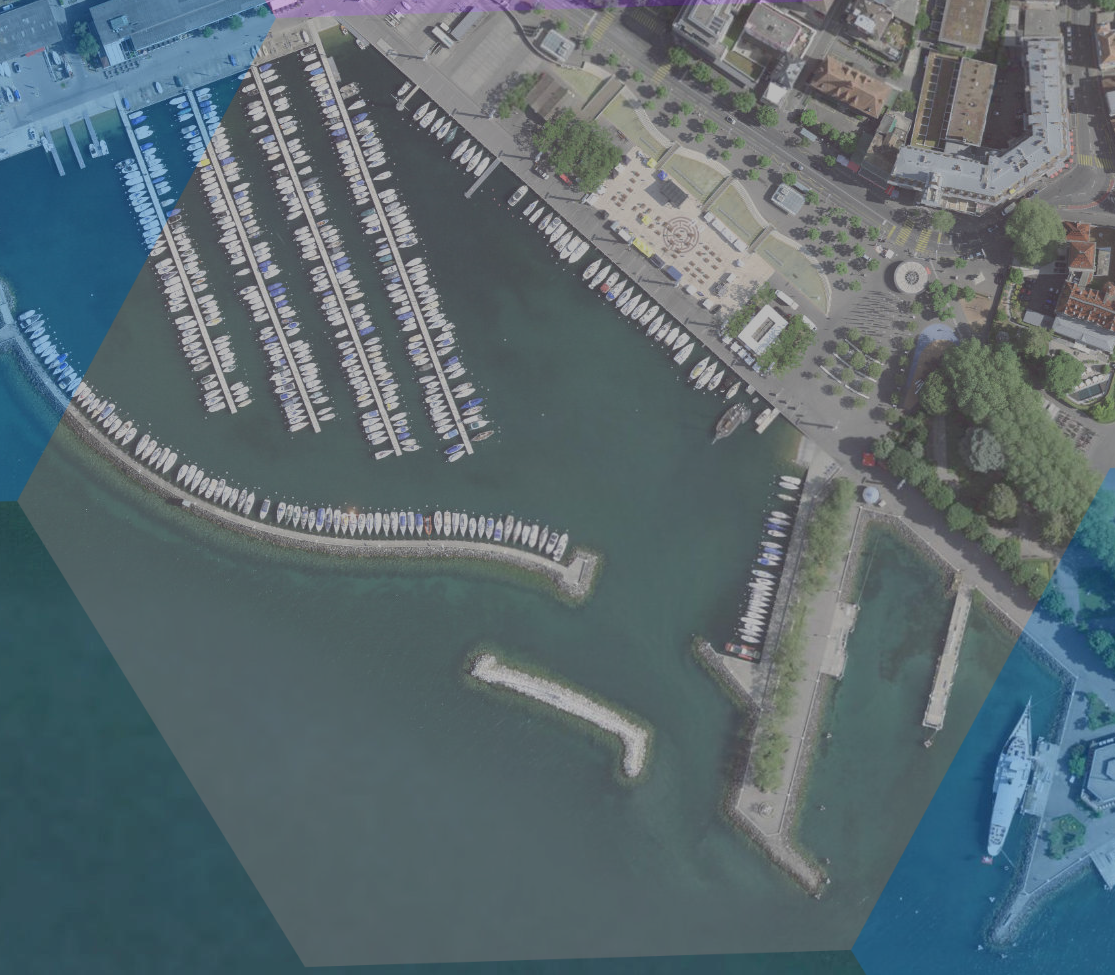}
        \caption{Satellite view of a second cell in Cluster~4 (Lausanne).}
        \label{fig:sat_view_2_lausanne}
    \end{subfigure}
    \caption{(a) and (b) Two distant cells from Cluster~4 in Lausanne. Despite their separation, each area features similar land-use characteristics, offering qualitative support for the clustering results.}
    \label{fig:lausanne_cluster4}
\end{figure}

\paragraph{Lausanne}
Two cells belonging to Cluster4 are presented in Figures\ref{fig:sat_view_1_lausanne} and \ref{fig:sat_view_2_lausanne}. These areas are characterized by a mix of natural elements—such as vegetation and proximity to water bodies—and major transportation hubs, including train and boat stations. Additionally, both neighborhoods show a notable presence of restaurants and hotels. Despite their spatial separation, the land-use patterns are strikingly similar, indicating that the clustering algorithm consistently identified comparable urban typologies. These qualitative observations align with the spatial distribution of Cluster4 in Figure\ref{fig:gmm_lausanne_5}, and further support the quantitative findings discussed in Section~\ref{subsec:quantitative_clusters}.

\paragraph{Philadelphia}

\begin{figure}[H]
    \centering
    \begin{subfigure}[b]{0.48\linewidth}
        \centering
        \includegraphics[width=\textwidth]{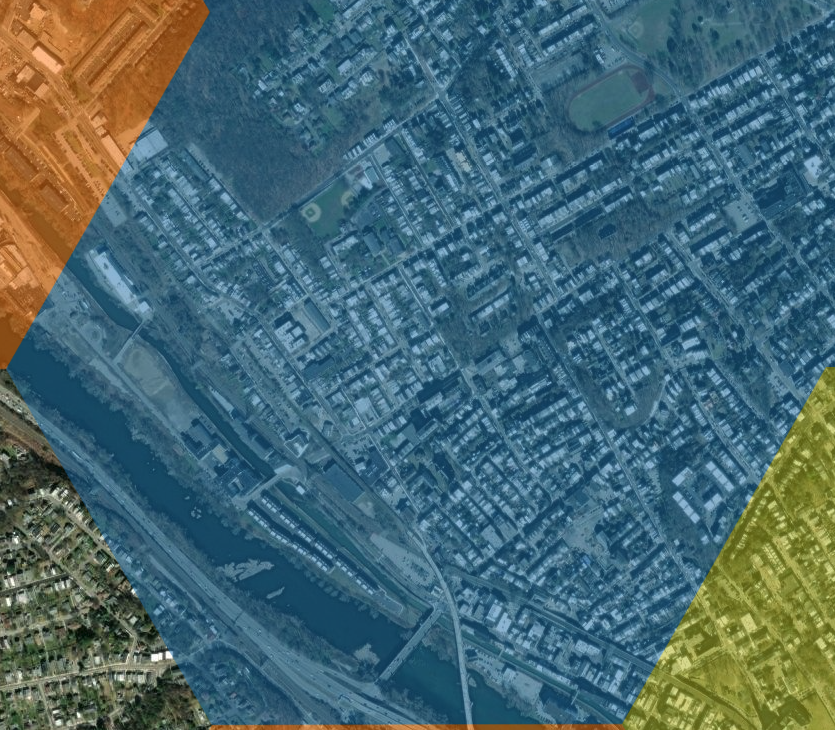}
        \caption{Satellite view of a first cell in Cluster~1 (Philadelphia).}
        \label{fig:sat_view_1_phily}
    \end{subfigure}
    \hfill
    \begin{subfigure}[b]{0.48\linewidth}
        \centering
        \includegraphics[width=\textwidth]{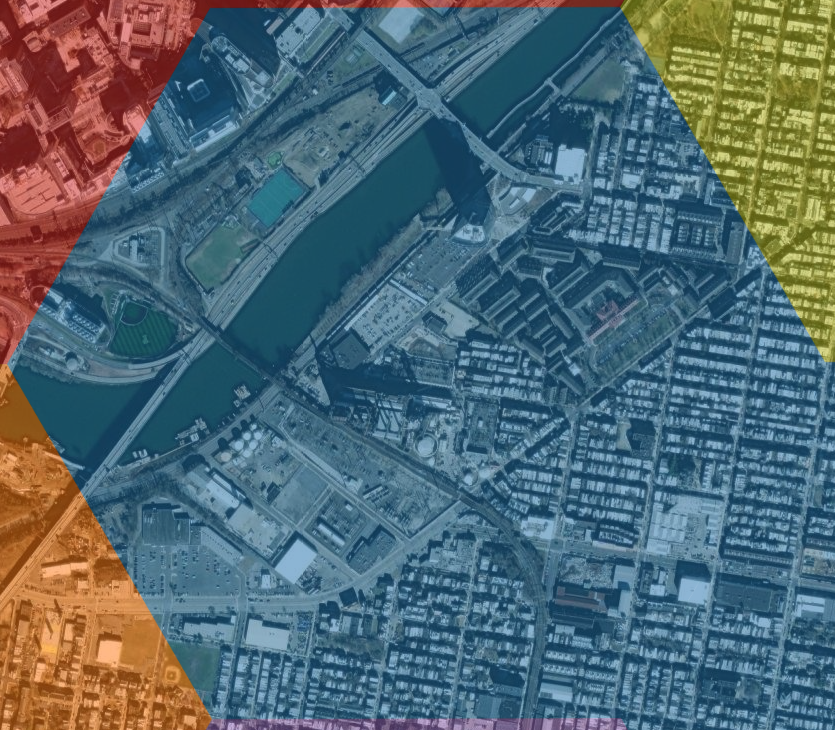}
        \caption{Satellite view of a second cell in Cluster~1 (Philadelphia).}
        \label{fig:sat_view_2_phily}
    \end{subfigure}
    \caption{(a) and (b) Two distant cells from Cluster~1 in Philadelphia. Despite their spatial separation, each location exhibits similarly dense residential patterns and comparable amenities, adding qualitative support to the clustering.}
    \label{fig:philly_cluster1}
\end{figure}

A comparable qualitative assessment is performed for Philadelphia by examining two spatially distant cells from Cluster1, shown in Figures\ref{fig:sat_view_1_phily} and \ref{fig:sat_view_2_phily}. Both areas are predominantly dense residential neighborhoods characterized by networks of small streets. Additionally, each location includes recreational facilities—such as baseball fields in one and tennis courts in the other—and natural features, including either a river or scattered tree cover. These shared land-use and infrastructural elements suggest that Cluster1 consistently captures a specific urban typology, despite local variations. This consistency supports the quantitative findings shown in Figure\ref{fig:gmm_philadelphia_7}, further validating the effectiveness of the clustering methodology.

\subsection{Comparative Analysis of Two Cities}
\label{subsec:two_cities}

A natural question emerging from this analysis is whether the urban typologies identified in one city are transferable to another. Specifically, do certain cluster profiles represent universal urban patterns, or are neighborhood characteristics inherently city-specific? To explore this, we merge the feature matrices of two distinct cities—Lausanne and Philadelphia—and apply a single clustering algorithm to the combined dataset. This approach allows us to assess the extent to which similar urban forms emerge across different geographic and cultural contexts.

\begin{figure}[H]
\centering
\includegraphics[width=0.8\textwidth]{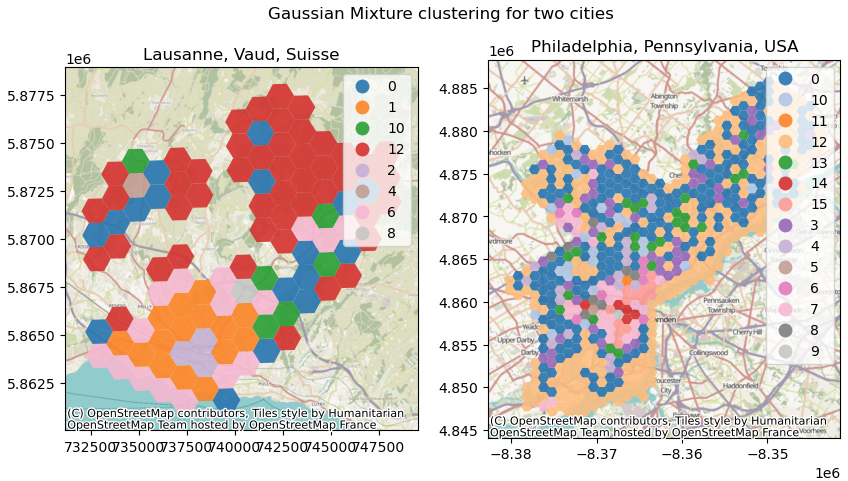}
\caption{Combined clustering result for Lausanne and Philadelphia with a uniform grid size.}
\label{fig:lausanne_phily}
\end{figure}

Figure~\ref{fig:lausanne_phily} demonstrates that some clusters are shared between Lausanne and Philadelphia when a uniform grid resolution is applied, suggesting the presence of comparable urban typologies across both cities. However, certain clusters—particularly Clusters 1 and 2, which are prominent in central Lausanne—are absent from the Philadelphia portion of the map. This discrepancy indicates that a single grid size may be insufficient to capture the varying spatial scales and urban structures of cities with distinct physical layouts and demographic profiles.

\begin{figure}[H]
\centering
\includegraphics[width=0.8\textwidth]{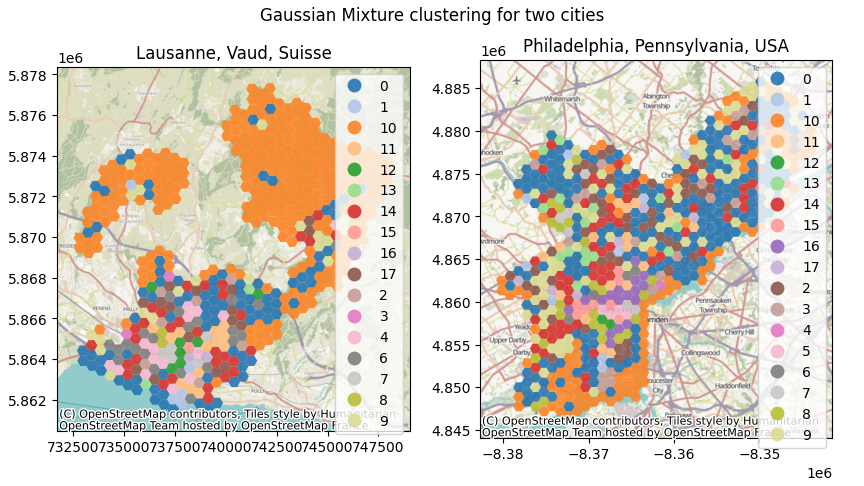}
\caption{Combined clustering result for Lausanne and Philadelphia with variable grid sizes.}
\label{fig:lausanne_phily_adapted}
\end{figure}

To address this limitation, we allow each city to be analyzed using an independently optimized grid size, as shown in Figure~\ref{fig:lausanne_phily_adapted}. With variable spatial resolutions, a greater number of shared clusters emerge between Lausanne and Philadelphia, including those corresponding to the urban cores of both cities. This result highlights the importance of adapting grid resolution to the specific spatial and morphological characteristics of each city. It reinforces the idea that appropriate scale selection is critical for accurately capturing comparable urban typologies across diverse geographic contexts.

\begin{figure}[H]
\centering
\includegraphics[width=1\textwidth]{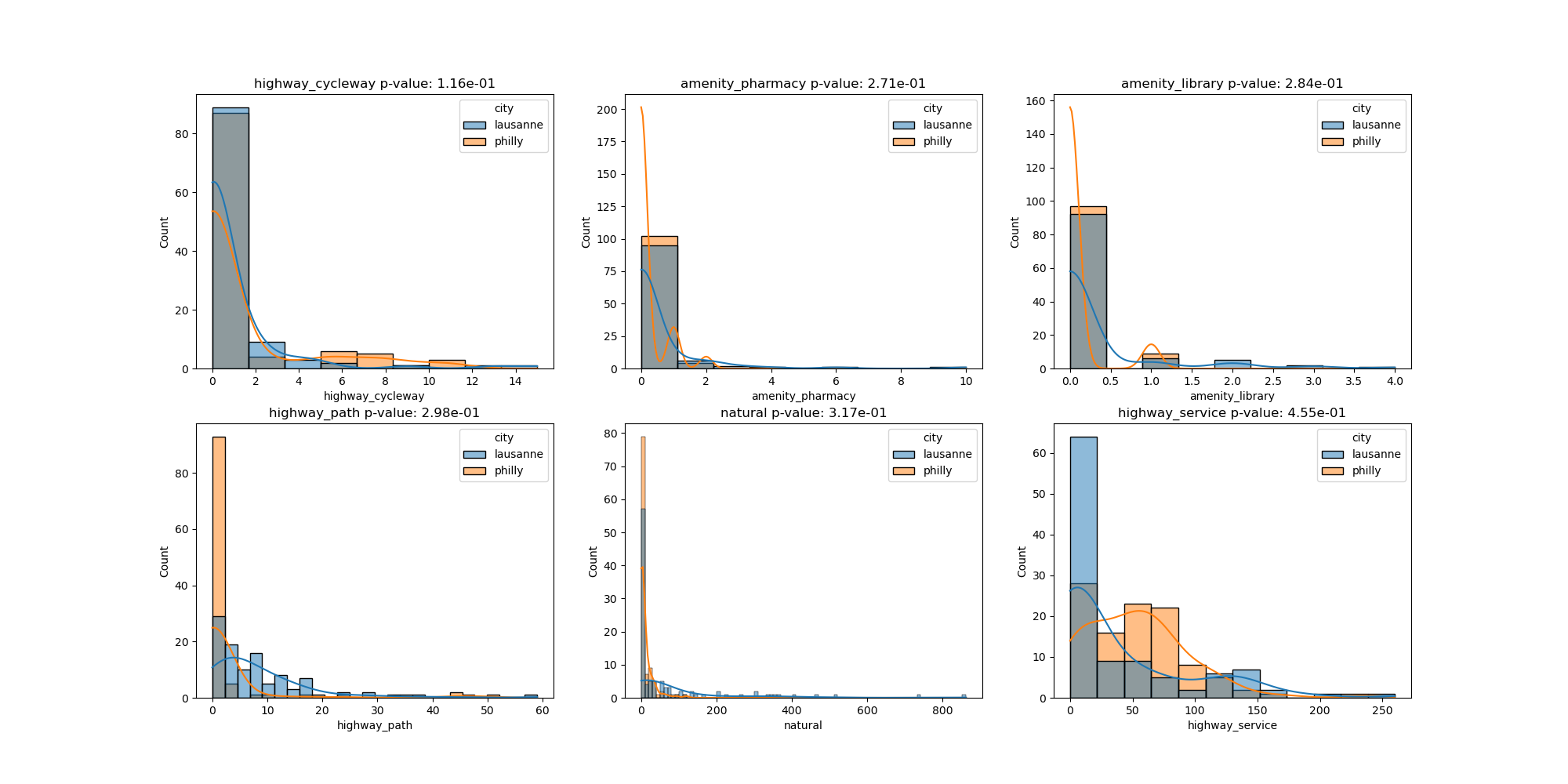}
\caption{Excerpt of the distribution of selected features for Lausanne and Philadelphia.}
\label{fig:features_distribution_laus_phil}
\end{figure}

Notably, the emergence of shared clusters between two cities located over 6,300 kilometers apart suggests that, when using a consistent set of features, comparable neighborhood typologies can be identified across distinct urban contexts. It is interesting to observe that, according to the data used for clustering, two BSUs within one city can be considered less alike than two BSUs located on two different continents This finding demonstrates the generalizability of the proposed clustering approach, which can be applied to multiple cities without requiring substantial methodological modifications. Further supporting this conclusion, Figure~\ref{fig:features_distribution_laus_phil} illustrates that many walkability-related features exhibit similar distributions in both Lausanne and Philadelphia. These parallels in feature distributions help explain the consistent appearance of certain cluster types across the two urban environments. Overall, these findings highlight that, when spatial resolution is appropriately tailored to each city's urban morphology, a unified clustering framework can effectively identify comparable neighborhood typologies across diverse urban environments.

\begin{figure}[H]
\centering
\includegraphics[width=13cm]{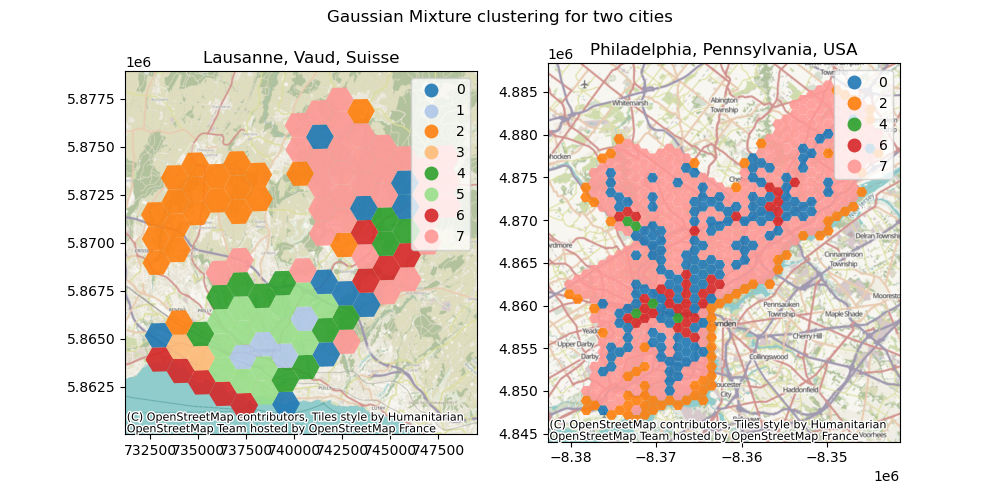}
\caption{Gaussian mixture clustering of the degree centrality of Lausanne and Philadelphia together.}\label{fig:networ_two_cities}
\end{figure}

Similarly, we extended the degree centrality-based clustering to the combined dataset of both Lausanne and Philadelphia to evaluate whether shared urban typologies based on network structure alone could be identified across cities. As shown in Figure~\ref{fig:networ_two_cities}, the clustering results reveal that while certain clusters are shared between the two cities, some typologies—particularly those characterized by very high street network degree—appear predominantly in Lausanne. This indicates that, although both cities contain similarly connected areas, Lausanne exhibits unique patterns of extremely dense street connectivity that are not as prevalent in Philadelphia.

\begin{figure}[H]
    \begin{subfigure}{.5\textwidth}
    \includegraphics[width=7cm]{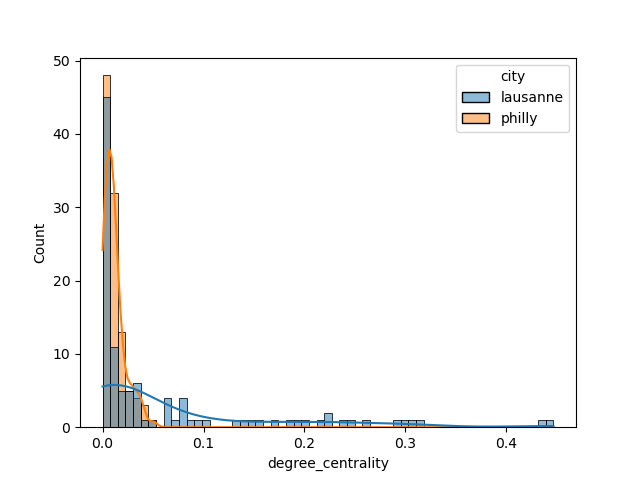}
    \caption{}\label{fig:network_two_cities_distrib}
    \end{subfigure}
    \begin{subfigure}{.5\textwidth}
    \includegraphics[width=6cm]{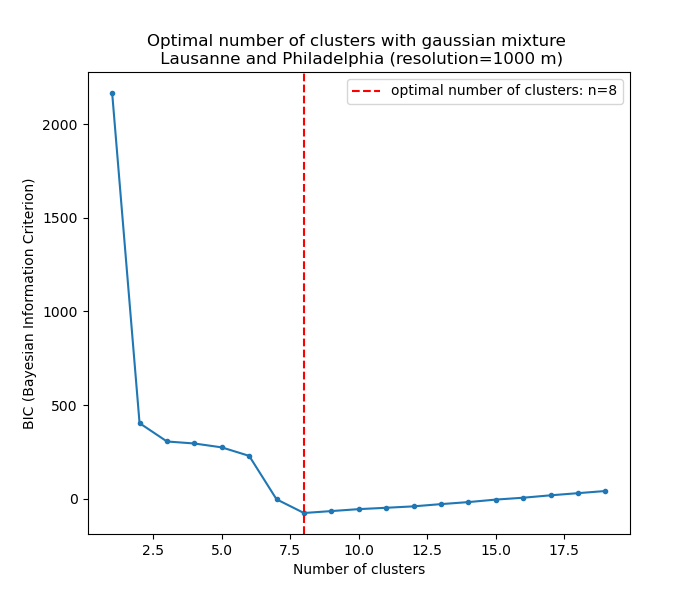}
    \caption{}\label{fig:network_two_cities_optim}
    \end{subfigure}
    \caption{(a) Distribution of degree centrality across cities. (b) Optimization of the number of clusters for two cities.}
\end{figure}

This observation is further supported by the comparison of degree centrality feature distributions between the two cities, illustrated in Figure~\ref{fig:network_two_cities_distrib}. The distributions highlight that Lausanne, as a compact and historically evolved European city, tends to have a more tightly knit street network compared to the broader, orthogonal grid system characteristic of Philadelphia.

An additional interesting insight emerges from the optimization analysis of the number of clusters using the Bayesian Information Criterion (BIC), shown in Figure~\ref{fig:network_two_cities_optim}. Here, we observe that the typologies present in Philadelphia are effectively captured with approximately five clusters, corresponding to a local minimum in the BIC curve. However, further increasing the number of clusters improves model fit predominantly for Lausanne, suggesting greater complexity and finer differentiation in its street network structure. This result highlights not only the morphological differences between the two cities but also reinforces the capacity of degree centrality alone to reveal multi-scalar patterns of urban form in a comparative context.

\section{Discussion} \label{sec:discussion}

This study introduced a scalable, data-driven framework for identifying and comparing urban form typologies across cities, using clustering techniques applied to open-source geospatial features. By leveraging a diverse set of features—from street types and public transport nodes to natural elements and amenities—we demonstrated that consistent, interpretable urban forms can be derived from cities as geographically and culturally distinct as Lausanne and Philadelphia. The findings offer important implications for urban planning practice, methodological approaches in spatial analysis, and comparative urbanism.

\subsection{Identifying Typologies Through Multi-Feature and Centrality-Based Clustering}

The use of a comprehensive feature set to define BSU allowed us to capture complex, multidimensional aspects of urban form. Clusters derived from this approach aligned well with known urban functions—such as commercial cores, residential zones, and multimodal nodes—validating both the clustering method and the selected features, supporting Hypothesis 1 (H1).

At the same time, the supplemental clustering using degree centrality as the sole feature offered a simplified, yet insightful lens into urban structure. This single-variable approach produced highly interpretable clusters tied directly to the density and configuration of street networks. Notably, in Philadelphia, the centrality-based clustering mirrored known gradients in urban form, from the dense orthogonal core to more irregular and sparse peripheries. When applied to both Lausanne and Philadelphia together, the degree centrality clusters revealed both shared and city-specific morphologies. For instance, high-degree centrality areas in Lausanne appeared unique and more prevalent than in Philadelphia, underscoring how historical development and topographic constraints shape network efficiency differently across contexts.

Together, these dual approaches—one holistic and feature-rich, the other minimal and structural—illustrate the versatility of clustering in urban analysis. Degree centrality, in particular, can serve as a computationally efficient proxy for urban form when data availability is limited, while still providing interpretable insights.

\subsection{Shared Patterns Across Cities and the Role of Scale}

One of the key contributions of this study lies in revealing that certain urban typologies emerge consistently across both cities, especially when grid resolution is optimized per city, supporting Hypothesis 4 (H4). In the joint clustering analysis, the application of a single grid resolution yielded fewer shared clusters due to spatial scale mismatches. However, when grid sizes were adapted to local morphology—450 meters for Lausanne and 1500 meters for Philadelphia—a broader set of shared cluster types emerged.

This finding supports Hypothesis 3 (H3) and reinforces the critical role of scale sensitivity in urban form analysis. Urban morphology is inherently multiscalar, and imposing uniform spatial units can distort the perception of shared structure. Allowing adaptive resolution ensures that clusters reflect local spatial logic while remaining comparable across contexts. This insight has methodological implications for global comparative studies and aligns with literature advocating for flexible spatial zoning in urban analytics \cite{lemoine2020global}.

\subsection{Functional Convergence and Transferability of Urban Typologies}

The emergence of shared clusters across cities with different planning histories, geographies, and cultures suggests a degree of functional convergence in urban environments. Central business districts, mixed-use neighborhoods, and transit-rich corridors appear to coalesce around similar patterns of amenities and infrastructure—even when separated by thousands of kilometers.

This outcome supports Hypothesis 2 (H2) and echoes broader research suggesting that, despite regional diversity, certain configurations of urban elements—such as multimodal access, high centrality, and rich amenity landscapes—support universal patterns of urban activity and walkability \cite{rapoport2013human, frank2001built}. From a policy perspective, this implies that design principles validated in one context may hold transferable value in others, especially when adapted to local scale and density.

\subsection{Clustering as a Diagnostic Tool for Planners}

The application of clustering, particularly through GMMs, offers a practical diagnostic tool for urban planners. By identifying areas that resemble highly walkable, amenity-rich typologies—or conversely, clusters characterized by poor connectivity and limited services—interventions can be strategically targeted. For instance, BSUs that fall into low centrality or low-feature-density clusters could be prioritized for infrastructure investment, green space development, or mobility enhancements.

Moreover, as the simplified degree centrality analysis shows, even a single feature can reveal structural inequalities or morphological constraints, helping planners visualize where urban fabric deviates from well-connected norms. This dual capability—broad typology identification and granular structural analysis—makes the framework adaptable to varying levels of data and planning capacity.

\subsection{Limitations and Future Directions}

While the study provides robust evidence for the utility of spatial clustering in urban form analysis, several limitations remain. The reliance on OSM data introduces potential inconsistencies in data completeness, especially for features like pedestrian crossings or green space, which may be underreported in certain areas. Additionally, qualitative elements such as perceived safety, lighting, or sidewalk quality—crucial to walkability—remain unrepresented in this data-driven framework.

Future work could expand the model by integrating sensor data, survey responses, or image-based assessments of the built environment. Furthermore, extending the approach to include more cities—especially those in the Global South or rapidly urbanizing regions—would test the generalizability of these findings. Lastly, incorporating temporal dynamics to explore how clusters evolve over time could provide valuable insights into the effects of policy interventions or urban development trends.

\section{Limitations}
\label{sec:limitations}

While the study presents a robust framework for clustering urban form based on walkability features, several limitations must be acknowledged. 

Firstly, the reliance on OpenStreetMap (OSM) data introduces inherent limitations related to data completeness and accuracy. Although OSM is a valuable resource for geospatial information, its data quality can vary significantly across different regions and features. Incomplete or outdated entries may lead to inaccuracies in the feature set, potentially affecting the clustering outcomes. Future studies could incorporate additional data sources or validation steps to mitigate these discrepancies.

Secondly, the selection of features, while comprehensive, may not encompass all dimensions of urban forms. Factors such as sidewalk quality, street lighting, safety, and pedestrian traffic are critical to walkability and subsequently urban forms but are challenging to quantify using OSM data alone. The exclusion of these qualitative aspects may result in an incomplete assessment of walkability. Integrating supplementary data sources, such as surveys or sensor-based measurements, could provide a more holistic understanding of pedestrian experiences.

Thirdly, the methodological approach, particularly the use of GMM, assumes that the underlying data distribution can be modeled as a mixture of Gaussian distributions. While GMMs offer flexibility, they may not capture more complex, non-Gaussian patterns present in urban data. Alternative clustering algorithms, such as Density-Based Spatial Clustering of Applications with Noise (DBSCAN) or hierarchical clustering, could be explored to address this limitation.

Additionally, the determination of optimal grid sizes and the number of clusters, although guided by quantitative metrics such as the Bayesian Information Criterion (BIC) and silhouette scores, remains somewhat subjective. Different optimization criteria or multi-scale approaches could yield varying results, potentially influencing the interpretation of walkability typologies. Future research could explore automated or adaptive grid-sizing techniques to enhance objectivity in this process.

Another limitation pertains to the generalizability of the findings. The study focuses on two cities with distinct characteristics, but the extent to which the results can be extrapolated to other urban contexts remains uncertain. Cities with vastly different urban forms, densities, or cultural practices may exhibit unique walkability profiles that are not captured by the current clustering framework. Expanding the analysis to include a more diverse set of cities would enhance the robustness and applicability of the methodology.

Lastly, the study does not incorporate direct feedback from residents or pedestrians, which is crucial for validating the perceived walkability of the identified clusters. Walkability is inherently subjective, influenced by individual preferences and experiences. Integrating qualitative data from surveys or participatory mapping could provide valuable insights into how well the clusters align with residents' perceptions of walkability, thereby strengthening the study's conclusions.

\section{Future Work}

Building on the current study, several promising directions for future work can be identified to enhance the framework’s robustness, scalability, and applicability to a wider range of urban contexts.

First, addressing data quality limitations remains a priority. Future research could integrate multiple open and proprietary data sources to supplement OpenStreetMap (OSM) information, including municipal datasets, satellite imagery, and sensor-based urban data. Such integration would allow for the inclusion of qualitative dimensions of urban form—such as sidewalk conditions, street lighting, or perceived safety—that are critical for understanding pedestrian experiences but are not fully captured by OSM alone.

Second, expanding the set of features and incorporating temporal dynamics could deepen insights into urban form evolution. Including time-sensitive variables, such as changes in land use, infrastructure upgrades, or seasonal variations in green space usage, would enable dynamic analyses of how urban typologies develop and transform over time.

Third, methodological advancements could further refine clustering performance. Future studies could explore non-parametric clustering techniques, such as DBSCAN or hierarchical clustering, to relax the Gaussian distribution assumptions inherent in GMMs. Additionally, adopting multi-scale clustering approaches, where multiple grid sizes are analyzed concurrently, could provide a richer, more nuanced understanding of urban spatial structures.

Fourth, expanding the comparative framework to a broader and more diverse set of cities—including cities from the Global South, rapidly urbanizing areas, and smaller towns—would allow a more comprehensive evaluation of the framework’s generalizability and reveal additional urban form typologies that may not be present in European or North American contexts.

Finally, integrating human-centered validation methods—such as surveys, participatory mapping workshops, or mobile sensing studies—would provide critical feedback on how well computationally identified clusters align with residents’ lived experiences. This human-in-the-loop approach would strengthen the framework’s utility for urban planning applications and foster more equitable, resident-informed urban interventions.

\section{Conclusion}
\label{sec:conclusion}
This study introduced a flexible, data-driven framework for analyzing and comparing urban forms across cities using clustering techniques based on spatial features extracted from OpenStreetMap. By applying the methodology to Lausanne and Philadelphia, we demonstrated that coherent and interpretable urban typologies can be identified both within and across geographically and culturally distinct cities. Our results highlight the critical role of scale in comparative urban analysis and show that even a single metric, such as degree centrality, can reveal meaningful structural patterns. While limitations exist in terms of data completeness and feature comprehensiveness, the framework’s adaptability across datasets, cities, and clustering approaches offers a strong foundation for future research. Our work contributes a reproducible method for diagnosing and understanding urban form, with potential applications in planning, policy-making, and the advancement of more walkable and sustainable cities.

\bibliographystyle{elsarticle-num}
\bibliography{references}

\begin{thebibliography}{10}
\expandafter\ifx\csname url\endcsname\relax
  \def\url#1{\texttt{#1}}\fi
\expandafter\ifx\csname urlprefix\endcsname\relax\def\urlprefix{URL }\fi
\expandafter\ifx\csname href\endcsname\relax
  \def\href#1#2{#2} \def\path#1{#1}\fi

\bibitem{haase2018global}
D.~Haase, B.~G{\"u}neralp, B.~Dahiya, X.~Bai, T.~Elmqvist, et~al., Global urbanization, The Urban Planet: Knowledge Towards Sustainable Cities 19 (2018) 326--339.

\bibitem{mouratidis2021urban}
K.~Mouratidis, Urban planning and quality of life: A review of pathways linking the built environment to subjective well-being, Cities 115 (2021) 103229.

\bibitem{kent2014three}
J.~L. Kent, S.~Thompson, The three domains of urban planning for health and well-being, Journal of planning literature 29~(3) (2014) 239--256.

\bibitem{tavakoli2025psycho}
A.~Tavakoli, I.~P. Douglas, H.~Y. Noh, J.~Hwang, S.~L. Billington, Psycho-behavioral responses to urban scenes: An exploration through eye-tracking, Cities 156 (2025) 105568.

\bibitem{panagopoulos2016urban}
T.~Panagopoulos, J.~A.~G. Duque, M.~B. Dan, Urban planning with respect to environmental quality and human well-being, Environmental pollution 208 (2016) 137--144.

\bibitem{barton2009land}
H.~Barton, Land use planning and health and well-being, Land use policy 26 (2009) S115--S123.

\bibitem{kropf2009aspects}
K.~Kropf, Aspects of urban form, Urban morphology 13~(2) (2009) 105--120.

\bibitem{dempsey2010elements}
N.~Dempsey, C.~Brown, S.~Raman, S.~Porta, M.~Jenks, C.~Jones, G.~Bramley, Elements of urban form, Dimensions of the sustainable city (2010) 21--51.

\bibitem{rapoport2013human}
A.~Rapoport, Human aspects of urban form: towards a man—environment approach to urban form and design, Elsevier, 2013.

\bibitem{frank2001built}
L.~D. Frank, P.~O. Engelke, The built environment and human activity patterns: exploring the impacts of urban form on public health, Journal of planning literature 16~(2) (2001) 202--218.

\bibitem{jackson2003relationship}
L.~E. Jackson, The relationship of urban design to human health and condition, Landscape and urban planning 64~(4) (2003) 191--200.

\bibitem{lemoine2020global}
R.~Lemoine-Rodr{\'\i}guez, L.~Inostroza, H.~Zepp, The global homogenization of urban form. an assessment of 194 cities across time, Landscape and Urban Planning 204 (2020) 103949.

\bibitem{schwanen2002urban}
T.~Schwanen, Urban form and commuting behaviour: a cross-european perspective, Tijdschrift voor economische en sociale geografie 93~(3) (2002) 336--343.

\bibitem{schwarz2010urban}
N.~Schwarz, Urban form revisited—selecting indicators for characterising european cities, Landscape and urban planning 96~(1) (2010) 29--47.

\bibitem{bosselmann2012urban}
P.~Bosselmann, Urban transformation: Understanding city form and design, Island Press, 2012.

\bibitem{boeing_measuring_2018}
G.~Boeing, Measuring the complexity of urban form and design, Urban Design International 23~(4) (2018) 281--292.

\bibitem{clifton2008quantitative}
K.~Clifton, R.~Ewing, G.-J. Knaap, Y.~Song, Quantitative analysis of urban form: a multidisciplinary review, Journal of Urbanism 1~(1) (2008) 17--45.

\bibitem{barros2017urban}
A.~P. Barros, L.~M. Mart{\'\i}nez, J.~M. Viegas, How urban form promotes walkability?, Transportation Research Procedia 27 (2017) 133--140.

\bibitem{ak2018urban}
A.~Ak, Urban form and walkability: The assessment of walkability capacity of ankara, Ph.D. thesis, Middle East Technical University (Turkey) (2018).

\bibitem{lee2018planning}
S.~Lee, J.~Koschinsky, E.~Talen, Planning tools for walkable neighborhoods: zoning, land use, and urban form, Journal of architectural and planning research (2018) 69--88.

\bibitem{chapman_smart_2021}
J.~Chapman, E.~Fox, W.~Bachman, L.~Frank, J.~Thomas, A.~R. Reyes, Smart {Location} {Database} {Technical} {Documentation} and {User} {Guide} ({Version} 3.0), Tech. rep., United States Environmental Protection Agency (Jun. 2021).

\bibitem{ewing_travel_2001}
R.~Ewing, R.~Cervero, \href{https://doi.org/10.3141/1780-10}{Travel and the {Built} {Environment}: {A} {Synthesis}}, Transportation Research Record 1780~(1) (2001) 87--114, publisher: SAGE Publications Inc.
\newblock \href {https://doi.org/10.3141/1780-10} {\path{doi:10.3141/1780-10}}.
\newline\urlprefix\url{https://doi.org/10.3141/1780-10}

\bibitem{cervero_travel_1997}
R.~Cervero, K.~Kockelman, \href{https://www.sciencedirect.com/science/article/pii/S1361920997000096}{Travel demand and the {3Ds}: {Density}, diversity, and design}, Transportation Research Part D: Transport and Environment 2~(3) (1997) 199--219.
\newblock \href {https://doi.org/10.1016/S1361-9209(97)00009-6} {\path{doi:10.1016/S1361-9209(97)00009-6}}.
\newline\urlprefix\url{https://www.sciencedirect.com/science/article/pii/S1361920997000096}

\bibitem{fonseca_built_2022}
F.~Fonseca, R.~, Paulo J.~G., C.~, Elisa, J.~, Mona, P.~, George, T.~, Simona, , R.~A.~R. Ramos, \href{https://doi.org/10.1080/15568318.2021.1914793}{Built environment attributes and their influence on walkability}, International Journal of Sustainable Transportation 16~(7) (2022) 660--679.
\newblock \href {https://doi.org/10.1080/15568318.2021.1914793} {\path{doi:10.1080/15568318.2021.1914793}}.
\newline\urlprefix\url{https://doi.org/10.1080/15568318.2021.1914793}

\bibitem{hajrasoulih2018urban}
A.~Hajrasoulih, V.~Del~Rio, J.~Francis, J.~Edmondson, et~al., Urban form and mental wellbeing: Scoping a theoretical framework for action, J. Urban Des. Ment. Health 5~(10) (2018).

\bibitem{jacobs2015indicators}
C.~Jacobs-Crisioni, M.~Kompil, C.~Baranzelli, C.~Lavalle, Indicators of urban form and sustainable urban transport, Joint Research Centre, European Commission: Ispra, Italy (2015) 1--36.

\bibitem{wu2024machine}
C.~Wu, J.~Wang, M.~Wang, M.-J. Kraak, Machine learning-based characterisation of urban morphology with the street pattern, Computers, Environment and Urban Systems 109 (2024) 102078.

\bibitem{yao2025decoding}
L.~Yao, K.~Peng, C.~Li, Y.~Huang, Decoding urban form through street networks: A nationwide analysis of chinese cities, Cities 158 (2025) 105711.

\bibitem{cao_analyzing_2013}
Z.~Cao, S.~Wang, G.~Forestier, A.~Puissant, C.~F. Eick, \href{https://dl.acm.org/doi/10.1145/2505821.2505827}{Analyzing the composition of cities using spatial clustering}, in: Proceedings of the 2nd {ACM} {SIGKDD} {International} {Workshop} on {Urban} {Computing}, {UrbComp} '13, Association for Computing Machinery, New York, NY, USA, 2013, pp. 1--8.
\newblock \href {https://doi.org/10.1145/2505821.2505827} {\path{doi:10.1145/2505821.2505827}}.
\newline\urlprefix\url{https://dl.acm.org/doi/10.1145/2505821.2505827}

\bibitem{gil_discovery_2012}
J.~Gil, J.~N. Beirão, N.~Montenegro, J.~P. Duarte, \href{https://journal.urbanform.org/index.php/jum/article/view/3966}{On the discovery of urban typologies: data mining the many dimensions of urban form}, Urban Morphology 16~(1) (2012) 27--40, number: 1.
\newblock \href {https://doi.org/10.51347/jum.v16i1.3966} {\path{doi:10.51347/jum.v16i1.3966}}.
\newline\urlprefix\url{https://journal.urbanform.org/index.php/jum/article/view/3966}

\bibitem{ng_clarans_2002}
R.~Ng, J.~Han, \href{https://ieeexplore.ieee.org/abstract/document/1033770}{{CLARANS}: a method for clustering objects for spatial data mining}, IEEE Transactions on Knowledge and Data Engineering 14~(5) (2002) 1003--1016.
\newblock \href {https://doi.org/10.1109/TKDE.2002.1033770} {\path{doi:10.1109/TKDE.2002.1033770}}.
\newline\urlprefix\url{https://ieeexplore.ieee.org/abstract/document/1033770}

\bibitem{boeing_urban_2019}
G.~Boeing, \href{https://doi.org/10.1007/s41109-019-0189-1}{Urban spatial order: street network orientation, configuration, and entropy}, Applied Network Science 4~(1) (2019) 67.
\newblock \href {https://doi.org/10.1007/s41109-019-0189-1} {\path{doi:10.1007/s41109-019-0189-1}}.
\newline\urlprefix\url{https://doi.org/10.1007/s41109-019-0189-1}

\bibitem{fan_quantifying_2024}
X.~Fan, R.~Chen, Y.~Lin, \href{https://dl.acm.org/doi/10.1145/3671127.3698187}{Quantifying the {Similarity} of {Urban} {Forms} through the {Building} {Morphology} {Complexity} {Index}}, in: Proceedings of the 11th {ACM} {International} {Conference} on {Systems} for {Energy}-{Efficient} {Buildings}, {Cities}, and {Transportation}, {BuildSys} '24, Association for Computing Machinery, New York, NY, USA, 2024, pp. 204--208.
\newblock \href {https://doi.org/10.1145/3671127.3698187} {\path{doi:10.1145/3671127.3698187}}.
\newline\urlprefix\url{https://dl.acm.org/doi/10.1145/3671127.3698187}

\bibitem{shahapure2020cluster}
K.~R. Shahapure, C.~Nicholas, Cluster quality analysis using silhouette score, in: 2020 IEEE 7th international conference on data science and advanced analytics (DSAA), IEEE, 2020, pp. 747--748.

\bibitem{zhao2015mixture}
J.~Zhao, L.~Jin, L.~Shi, Mixture model selection via hierarchical bic, Computational Statistics \& Data Analysis 88 (2015) 139--153.

\end{thebibliography}

\section{Acknowledgments}

The authors would like to thank the Villanova University College of Engineering Faculty Career Development Award for supporting this work. 

\section{Supplementary results}
\label{appendix:supplementary_results}

\subsection{Example of complete set of results for Philadelphia}
\begin{figure}[H]
    \begin{subfigure}{.5\textwidth}
    \includegraphics[width=7cm]{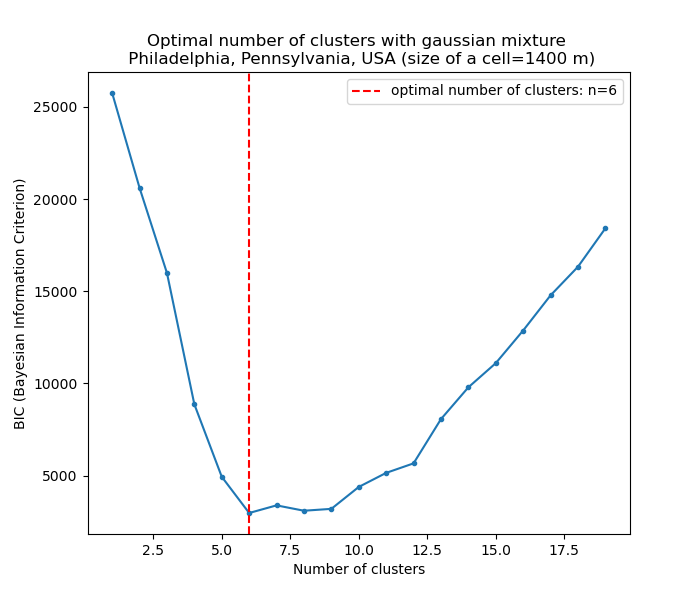}
    \caption{}
    \end{subfigure}
    \begin{subfigure}{.5\textwidth}
    \includegraphics[width=8cm]{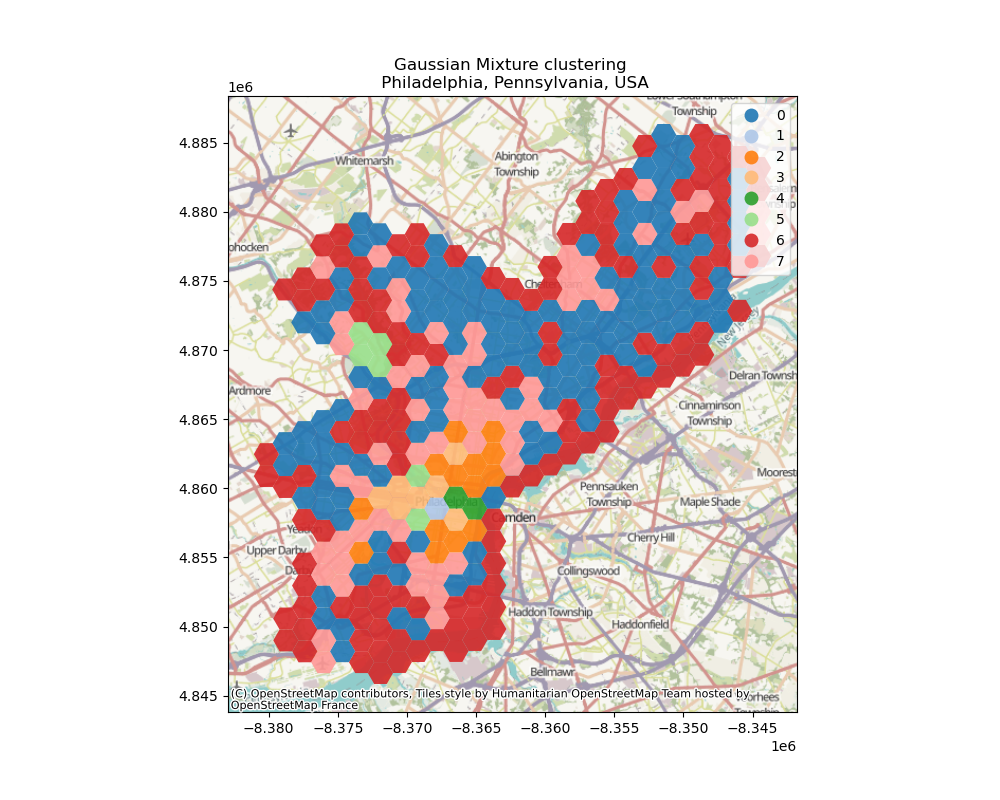}
    \caption{}
    \end{subfigure}
    \caption{(a) Optimization of the number of clusters. (b) Visualization of the result of clustering.}
\end{figure}

\begin{figure}[H]
\centering
\includegraphics[width=13cm]{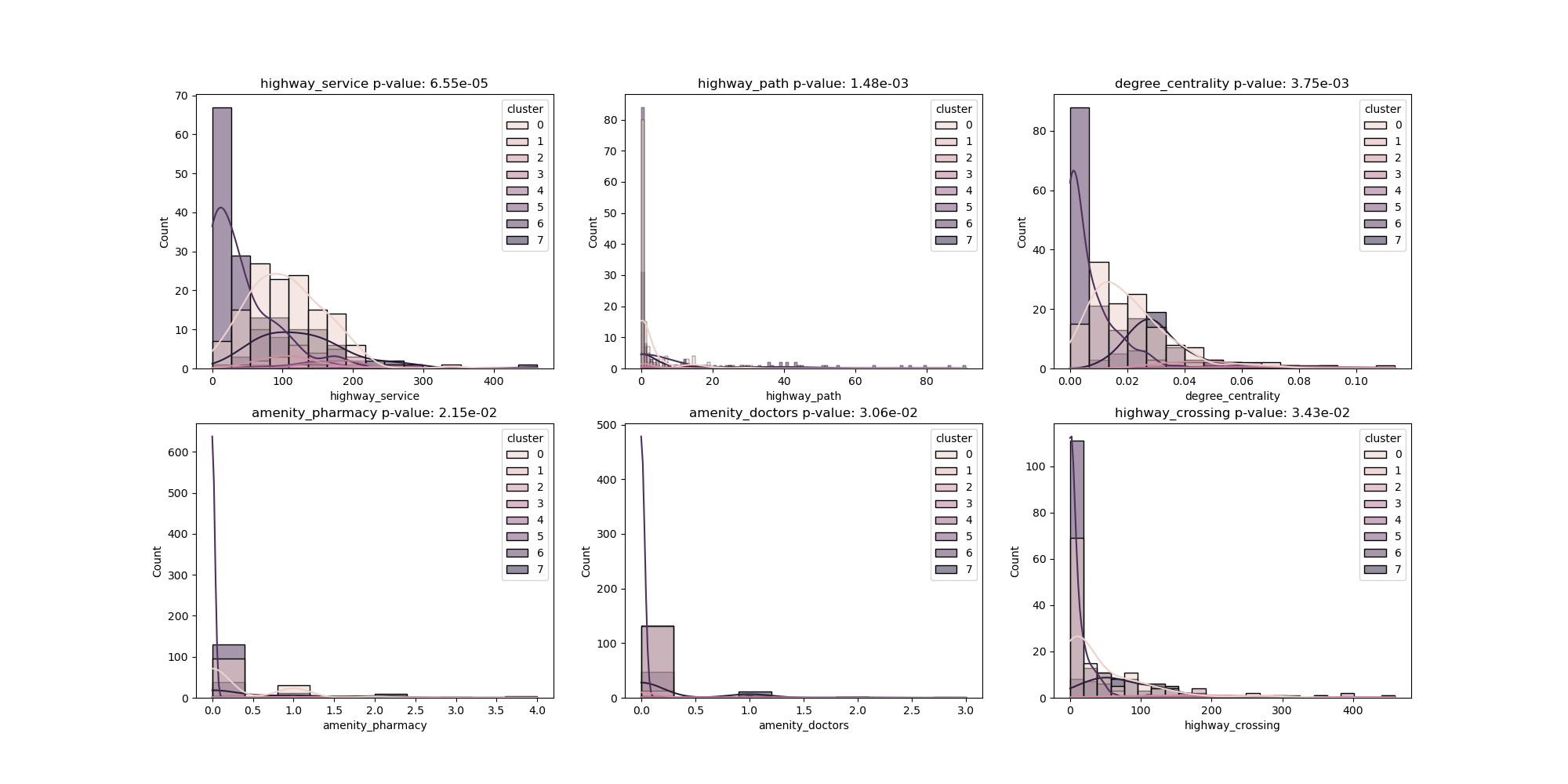}
\caption{Distribution of the values of features over the different clusters.}
\end{figure}

\begin{figure}[H]
\centering
\includegraphics[width=7cm]{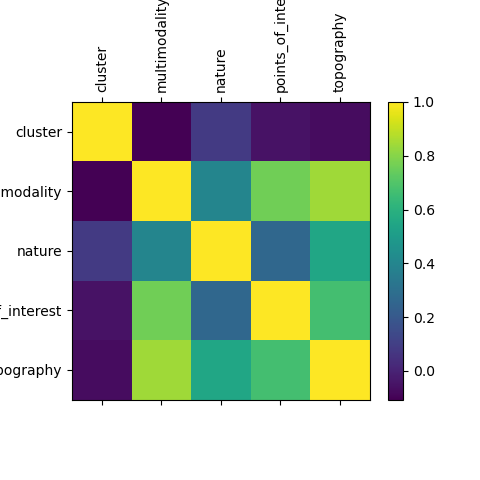}
\caption{Correlation matrix of the features grouped by categories.}
\end{figure}

\begin{figure}[H]
\centering
\includegraphics[width=13cm]{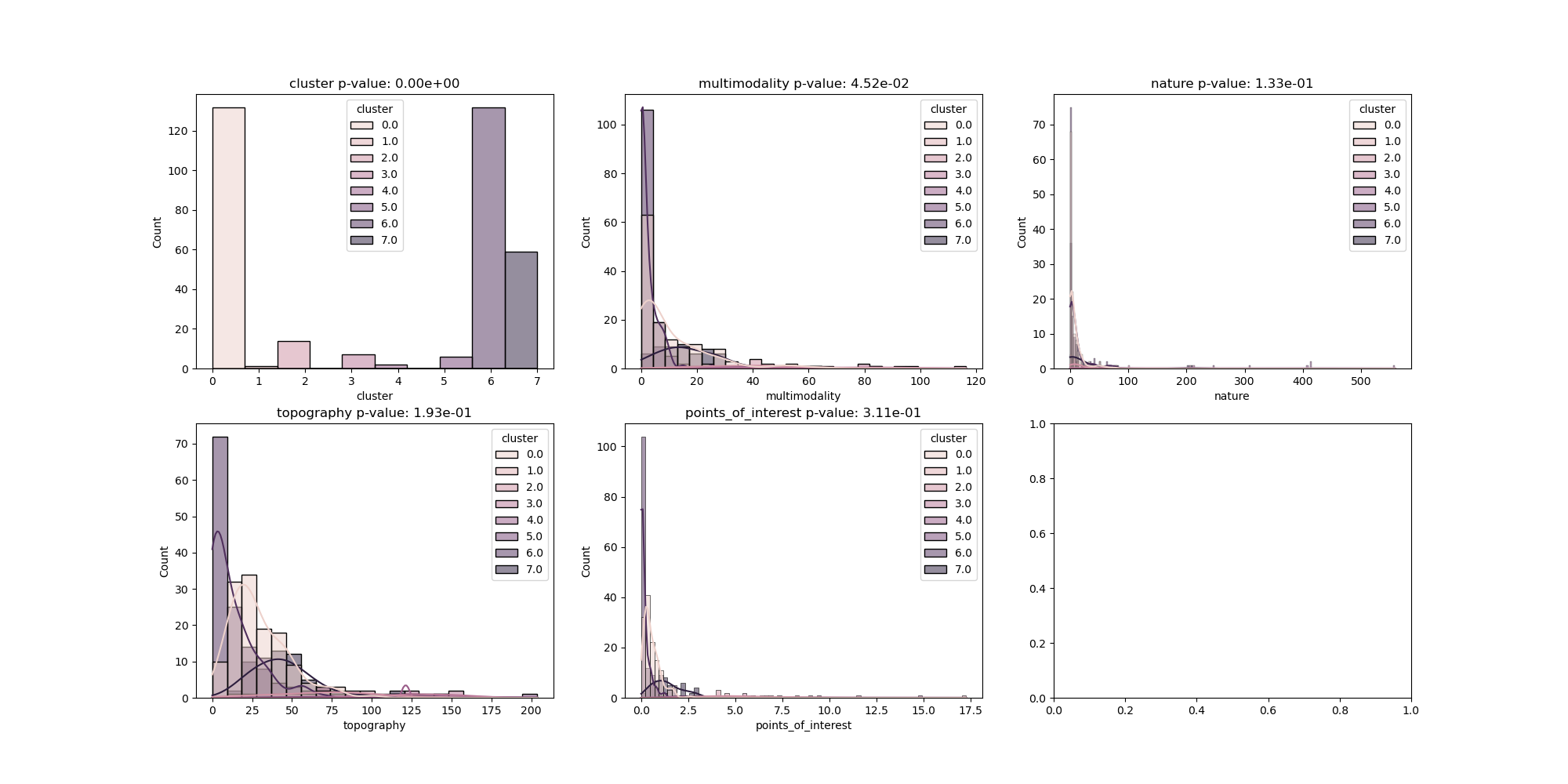}
\caption{Distribution of the values of features grouped by categories over the different clusters.}
\end{figure}

\subsection{Example of complete set of results for Lausanne}
\begin{figure}[H]
    \begin{subfigure}{.5\textwidth}
    \includegraphics[width=7cm]{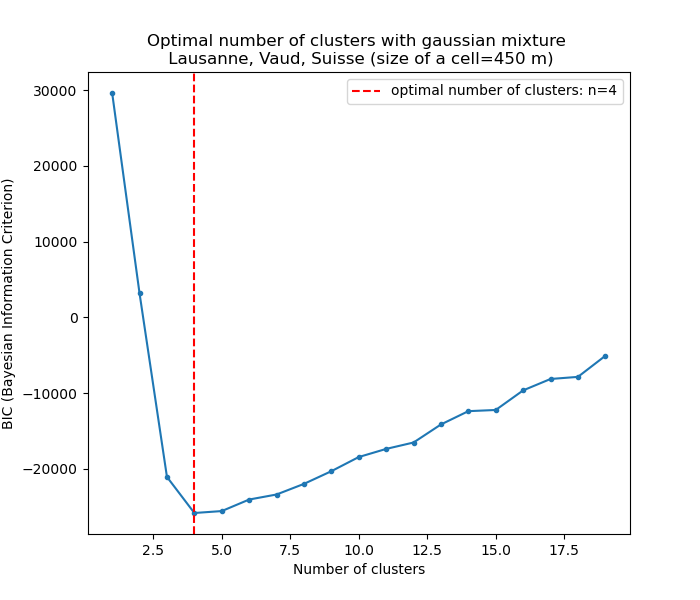}
    \caption{}
    \end{subfigure}
    \begin{subfigure}{.5\textwidth}
    \includegraphics[width=8cm]{Suisse_450_clusters.png}
    \caption{}
    \end{subfigure}
    \caption{(a) Optimization of the number of clusters. (b) Visualization of the result of clustering.}
\end{figure}

\begin{figure}[H]
\centering
\includegraphics[width=13cm]{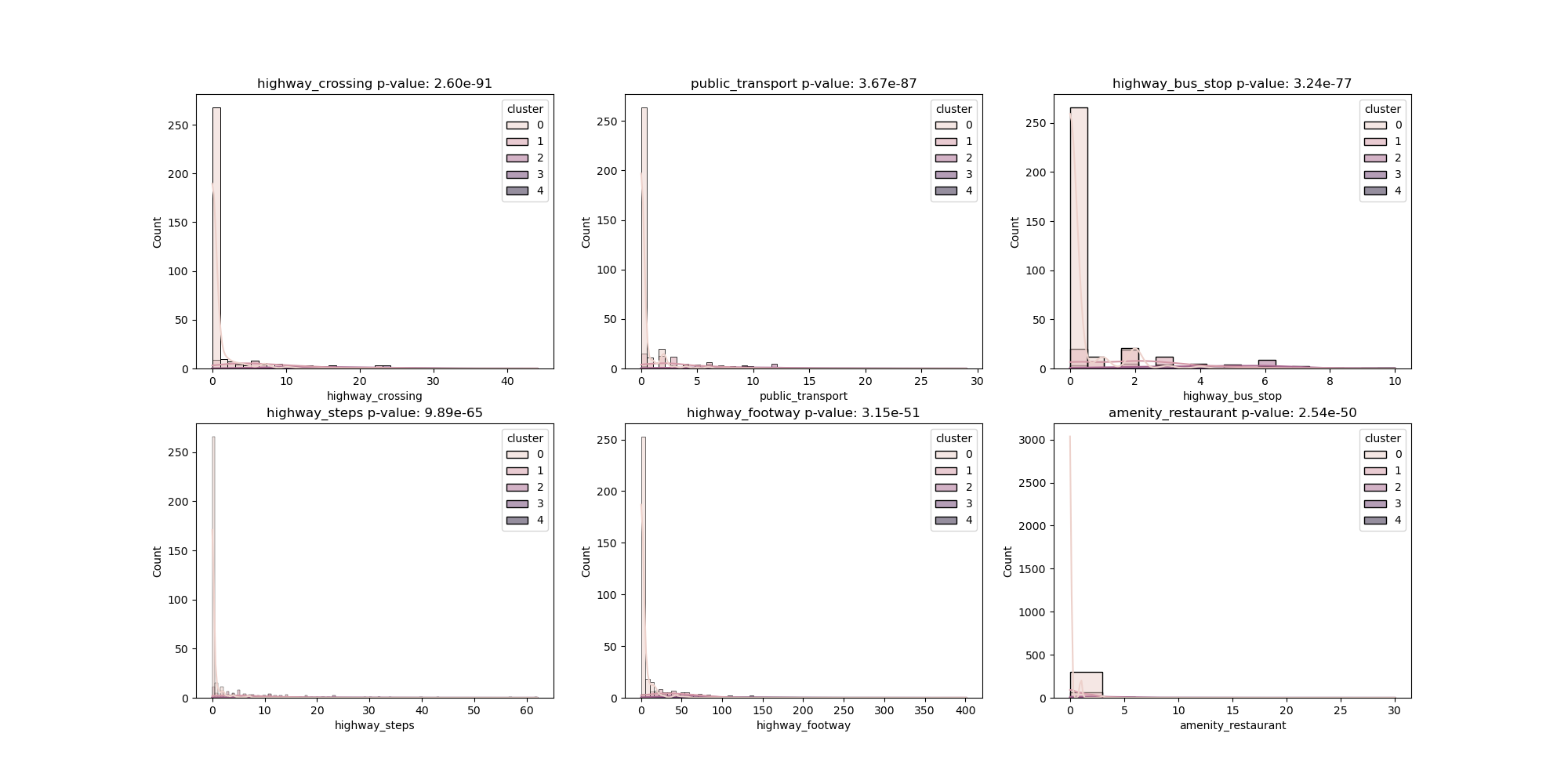}
\caption{Distribution of the values of features over the different clusters.}
\end{figure}

\begin{figure}[H]
\centering
\includegraphics[width=7cm]{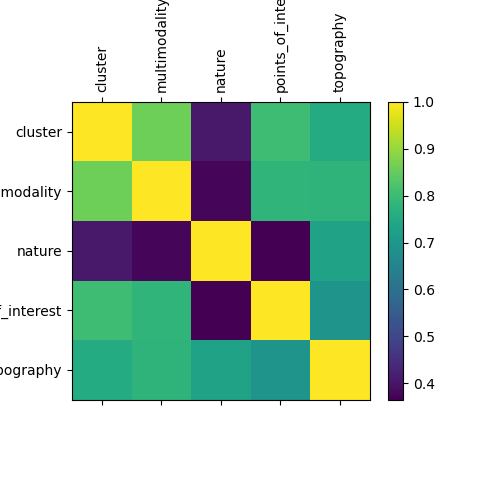}
\caption{Correlation matrix of the features grouped by categories.}
\end{figure}

\begin{figure}[H]
\centering
\includegraphics[width=13cm]{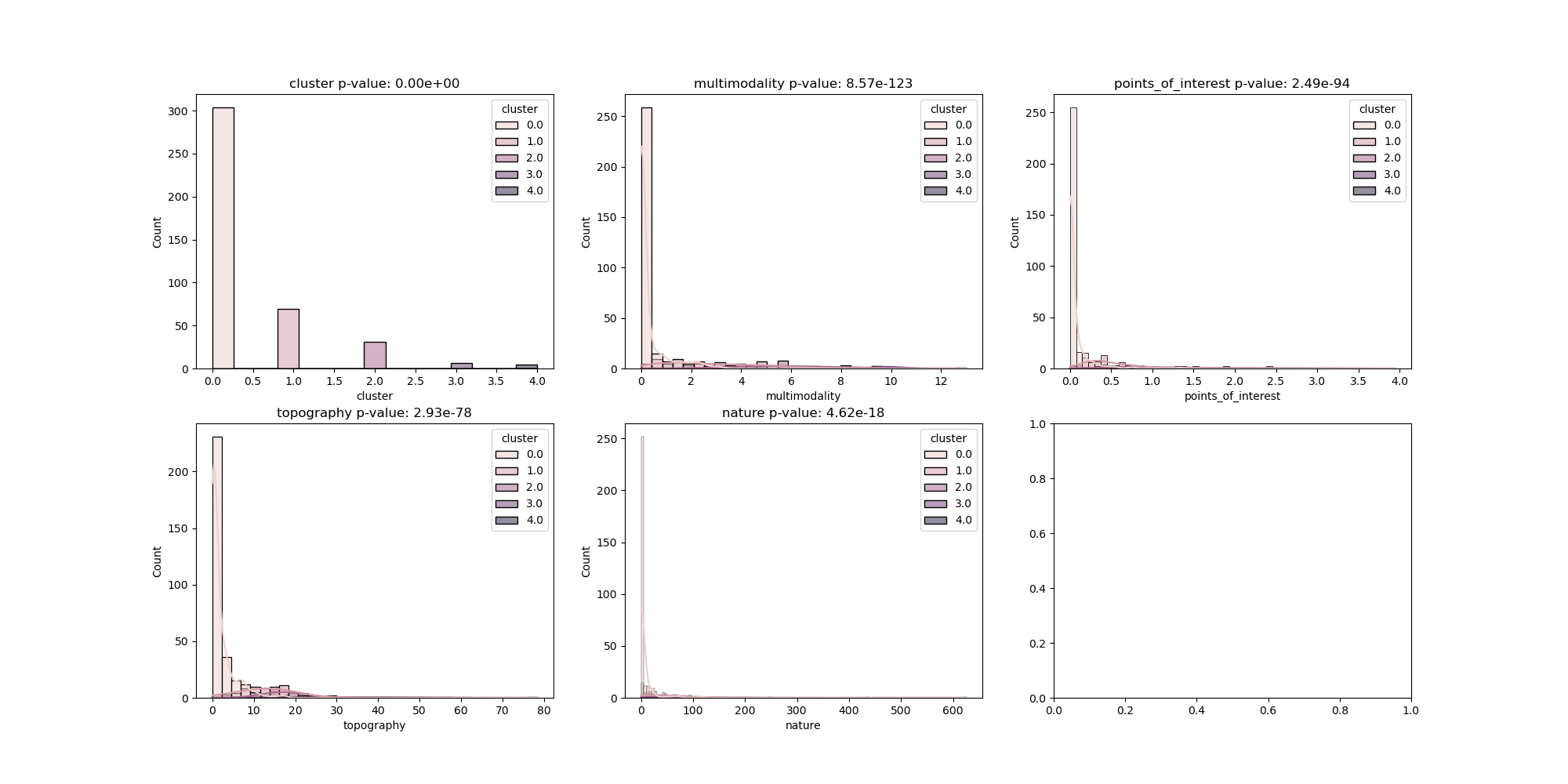}
\caption{Distribution of the values of features grouped by categories over the different clusters. Probably exhibits a clustering too much influenced by natural elements.}
\end{figure}

\end{document}